\documentclass[sigconf,balance=false]{acmart}
\usepackage{popets}

\usepackage{wrapfig}
\usepackage{subfiles}
\usepackage[utf8]{inputenc}
\usepackage{enumitem}
\setlist[itemize]{noitemsep, nolistsep, leftmargin=*}
\setlist[enumerate]{noitemsep, nolistsep, leftmargin=*}
\usepackage{bm}
\usepackage{adjustbox}
\usepackage{tabularx}
\usepackage{graphicx,subcaption}
\usepackage[title]{appendix}
\usepackage{wrapfig,lipsum,booktabs}

\setcopyright{popets}
\copyrightyear{YYYY}

\acmYear{YYYY}
\acmVolume{YYYY}
\acmNumber{X}
\acmDOI{XXXXXXX.XXXXXXX}
\acmISBN{}
\acmConference{Proceedings on Privacy Enhancing Technologies}
\newif\ifcomments
\commentstrue
\ifcomments
    \newcommand{\ys}[1]{{\textcolor{red}{-#1-}}}
    \newcommand{\zl}[1]{{\textcolor{blue}{#1}}}
\else
    \newcommand{\ys}[1]{}
    \newcommand{\zl}[1]{}
\fi

\begin{document}

\title[RPKI-based Tor relay selection]{RPKI-Based Location-Unaware \\ Tor Guard Relay Selection Algorithms}

\author{Zhifan Lu}
\affiliation{%
  \institution{University of Virginia}
  \city{Charlottesville} 
  \state{Virginia} 
  \country{USA} 
}
\email{zl2da@virginia.edu}

\author{Siyang Sun}
\affiliation{%
  \institution{University of Virginia}
  \city{Charlottesville}
  \state{Virginia}
  \country{USA}}
\email{ss6cuf@virginia.edu}

\author{Yixin Sun}
\affiliation{%
  \institution{University of Virginia}
  \city{Charlottesville}
  \state{Virginia}
  \country{USA}}
\email{ys3kz@virginia.edu}

\renewcommand{\shortauthors}{Lu et al.}

\begin{abstract}

Tor is a well-known anonymous communication tool, used by people with various privacy and security needs. Prior works have exploited routing attacks to observe Tor traffic and deanonymize Tor users. Subsequently, location-aware relay selection algorithms have been proposed to defend against such attacks on Tor. However, location-aware relay selection algorithms are known to be vulnerable to information leakage on client locations and guard placement attacks. 
Can we design a new location-unaware approach to relay selection while achieving the similar goal of defending against routing attacks?

Towards this end, we leverage the Resource Public Key Infrastructure (RPKI) in designing new guard relay selection algorithms. 
We develop a lightweight Discount Selection algorithm by only incorporating Route Origin Authorization (ROA) information, and a more secure Matching Selection algorithm by incorporating both ROA and Route Origin Validation (ROV) information. 
Our evaluation results show an increase in the number of ROA-ROV matched client-relay pairs using our Matching Selection algorithm, reaching 48.47\% with minimal performance overhead through custom Shadow simulations and benchmarking.


\end{abstract}


\maketitle

\section{Introduction}





The Tor network~\cite{Dingledine_Mathewson_Syverson2004} has been the most widely-used anonymous communication system to protect user identity in online communications. Tor is a low-latency system that does not obfuscate packet sizes and timings by default. 
However, Tor's low-latency nature makes it vulnerable to traffic analysis, where an adversary who can observe Tor traffic can deanonymize the users based on traffic patterns. 
The most well known attacks are traffic correlation attacks \cite{Zhu_Fu_Graham_Bettati_Zhao2010} where an adversary observes both ends of the communication path (i.e., between the Tor client and the entry/guard relay, and
between the exit relay and the destination server) and correlates the traffic, and website fingerprinting attacks \cite{Cherubin_Jansen_Troncoso2022, Juarez_Afroz_Acar_Diaz_Greenstadt2014} where an adversary observes client-side Tor traffic (i.e., between the client and the entry/guard relay) and matches it to the traffic patterns of known websites. 


Network-level adversaries, i.e., Autonomous
Systems (ASes), that lie on the communication paths are at a powerful position to observe Tor traffic and perform traffic analysis. 
To defend against network-level adversaries, many \emph{location-aware} relay selection algorithms have been proposed \cite{Barton_Wright2016, Edman_Syverson2009, Hanley_Sun_Wagh_Mittal2019, Kohls_Jansen_Rupprecht_Holz_Poepper2019, Sun_Edmundson_Feamster_Chiang_Mittal2017, Nithyanand_Starov_Zair_Gill_Schapira_2015, Li_Herwig_Levin2017, Rochet_Wails_Johnson_Mittal_Pereira2020}, which perform relay selection based on client locations to avoid or lower the risk of having an adversary on the path. 
In particular, Counter-RAPTOR \cite{Sun_Edmundson_Feamster_Chiang_Mittal2017} defends against the strongest network-level adversaries who can launch active routing attacks to hijack the client-guard connection, by strategically selecting guard relays to minimize the possibility of a successful hijack. 

However, similar to all other \emph{location-aware} relay selection algorithms, Counter-RAPTOR suffers from information leakage of client locations due to the differing relay weights that are tied to client locations. While there exist several works that provide an upper bound on the amount of information leakage \cite{Hanley_Sun_Wagh_Mittal2019, Rochet_Wails_Johnson_Mittal_Pereira2020}, they unavoidably face the tradeoff between limiting information leakage and achieving the main location-aware relay selection goal. 

Can we avoid this tradeoff by developing a location-agnostic relay selection algorithm to defend against the active network-level adversaries, similar to Counter-RAPTOR? 
We answer this question by utilizing recent advances in the deployment of Resource Public Key Infrastructure (RPKI). 
In a nutshell, an AS publishes a cryptographically-signed Route Origin Authorization (ROA) object that attests its authorization to announce an IP prefix. ASes also perform Route Origin Validation (ROV) on routing announcements, and drop the invalid routes that violate any existing ROAs (i.e., the route is announced by an AS who does not have a valid ROA record for the IP prefix).
Our intuition is that clients should favor relays with valid ROAs for the prefix that covers the relay IP, which naturally protects clients against active routing attacks that hijack the Tor relay IP. 
This is made possible by the rapid increase in ROA and ROV deployment by ASes in recent years.


More specifically, we focus on guard relay selection given the unique position of the guard relay that enables it to observe client traffic for an extended period of time. 
We design two versions of the RPKI-based guard relay selection algorithms: (1) a lightweight \emph{discount algorithm} that only considers whether the guard relay's IP is protected by a valid ROA, and ``discounts" the selection probability if the relay is not ROA-protected; (2) a \emph{matching algorithm} that aims to fully utilize the benefit of RPKI by increasing the probability of a ROA-protected client choosing a relay in a ROV-enforcing AS (and vice versa). The intuition is that ROA protection is the most effective when ASes perform ROV to drop any invalid routes. 

To evaluate the feasibility of our approach, we first perform a measurement study on the ROA coverage of Tor relays from 2021 to 2024. We then integrate performance considerations in relay selection by building upon the linear-programming framework used in CLAPS \cite{Rochet_Wails_Johnson_Mittal_Pereira2020}. We evaluate the security improvement in terms of protection against route origin hijacks through simulations, as well as the performance through a large-scale simulation via the Shadow simulator \cite{Jansen_Hopper2011}. 
Our key results are:
\begin{itemize}
\item The ROA coverage of Tor relays has been steadily growing. In particular, the percentage of ROA-protected guard relays grows from 47.20\% in Jan 2021 to 71.43\% in May 2024. Such considerable ROA coverage is the enabler for our RPKI-based relay selection approach. 
\item In the lightweight discount algorithm, the discount value is affected by the available bandwidth of ROA-protected relays, as well as the current load in the network. For example, when the network load is 80\% of the maximum load (i.e., saturating all available guard relay bandwidth), using consensuses from May 2024, a discount value of 0.5 provides ROA-protected guard relays to 90.2\% of clients (18.8\% improvement compared to Vanilla Tor) without saturating any single relay.
\item The matching algorithm significantly increases the number of client/guard pairs that can benefit from the full ROA/ROV protection. Using a combination of $l$ = 0.8, $d_{1}$ = 0.9,$d_{2}$ = 0.7 and $B$ = 1.5, 48.47\% of all client-relay pairs are matched with both ROA and ROV protection.
\item The Shadow simulation shows no visible difference in load time for discount algorithm; for matching algorithm, the overhead is minimal for small page loads (e.g., 10MB), while more significant for large size data transfers (e.g., 100MB).
\end{itemize}

Unlike prior works, our relay selection algorithms are not location-dependent. While there may exist some information leakage in the matching algorithm where the ROA/ROV status of the client AS may be learned, it is very challenging to pinpoint the exact AS only based on the ROA/ROV status. The potential leakage is significantly less even compared to the state-of-the-art CLAPS \cite{Rochet_Wails_Johnson_Mittal_Pereira2020}. 
Furthermore, unlike Counter-RAPTOR that aims to increase \emph{probabilistic} resilience to attacks, our usage of RPKI provides attack resilience via \emph{deterministic} validation of route origin. 

\textbf{Code release.} Our have open sourced our code at: \url{https://github.com/z-lu2017/TOR-RPKI}.

\section{Background and Motivation}

The Tor network aims to provide anonymous communications over the Internet. To achieve its goal, Tor implements onion routing and routes all traffic through a sequence of three relays, i.e., the entry/guard relay, the middle relay and the exit relay, before reaching the destination server. 
This prevents the eavesdropper from linking the client and the server together. 

Currently, the default path selection algorithm to choose relays is based on relay flags (e.g., relays with the "guard" flag can be chosen as the guard relay) and its weight in the network consensus for performance and load balancing purpose.


\subsection{AS-level Adversaries and Location-aware Relay Selection}
Tor is known to be vulnerable to traffic analysis that deanonymizes users. 
Many works have examined AS-level adversaries who are on path to observe Tor communications \cite{Sun_Edmundson_Vanbever_Li_Rexford_Chiang_Mittal2015, Akhoondi_Yu_Madhyastha2012, Edman_Syverson2009, Nithyanand_Starov_Zair_Gill_Schapira_2015, Johnson_Jansen_Jaggard_Feigenbaum_Syverson2017}, the most powerful of which is RAPTOR attack \cite{Sun_Edmundson_Vanbever_Li_Rexford_Chiang_Mittal2015} where the adversary manipulates Internet routing to announce the IP prefix of a guard relay and route Tor client traffic to the adversary. 
On the other hand, many client location-aware relay selections have been proposed, serving various purposes such as minimizing vulnerability to AS-level adversaries \cite{Edman_Syverson2009, Hanley_Sun_Wagh_Mittal2019, Kohls_Jansen_Rupprecht_Holz_Poepper2019, Sun_Edmundson_Feamster_Chiang_Mittal2017, Nithyanand_Starov_Zair_Gill_Schapira_2015, Li_Herwig_Levin2017} or communication latency \cite{Barton_Wright2016, Snader_Borisov2008, Wang_Bauer_Forero_Goldberg_Keromytis2012, Sherr_Blaze_Loo_Goldberg_Atallah2009}. In essence, the probability of selecting a given relay differs across clients and is determined based on client location. 
Such location-aware relay selection algorithms are known to be vulnerable to information leakage and guard placement attacks. 


\textbf{Client location leakage.} 
Location-aware path selection algorithms can inadvertently leak client location information. Because clients prefer relays with high bandwidth that are close in physical or topological distance, if an adversary can observe which relays a client connects to, then she/he can obtain information about the client's location. For example, \cite{Hanley_Sun_Wagh_Mittal2019} shows an adversary can successfully guess client AS location under 10 attempts after observing 5 relay selections, and under 5 attempts if 15 selections are observed. \cite{Johnson_Jansen_Jaggard_Feigenbaum_Syverson2017} demonstrates a chosen-destination attack by forcing client go through various destinations to reveal client guard relay, with guard relays being found 94\% of the times after 300 visits. \cite{Wails_Sun_Johnson_Chiang_Mittal2018} analyzes information leakage in DeNASA \cite{Barton_Wright2016} and finds that after observing three guard selections, median entropy for all ASes drop to 1 bit, making it easy for adversaries to guess client location using posterior distribution of guard-selection for all client locations.

\textbf{Guard placement attack}. Guard placement attack happens when an adversary places malicious guard relays in the Tor network with the goal of maximizing the probability that a client entering the network will choose one of the malicious guard relays. 
Such attack is particularly effective against location-aware relay selection algorithms, where the adversary can strategically place relays in certain locations to maximize the probability of being selected~\cite{Wan_Johnson_Wails_Wagh_Mittal2019}. 

In summary, while location-aware relay selection can provide many benefits, it unavoidably creates new vulnerabilities as described above.

\subsection{Resource Public Key Infrastructure (RPKI)}
Resource Public Key Infrastructure (RPKI) is a public key infrastructure framework designed as a security improvement to the current routing protocol BGP. 
RPKI provides a database of cryptographically-signed IP address ownership that can be
used to identify and drop route announcements from unauthorized ASes \cite{ROA_Database}.


\textbf{Route Origin Authorization (ROA)} is a cryptographically signed record stored in RPKI that includes (AS, prefix) pairs, indicating that the AS is authorized to announce the prefix. The signature can be cryptographically verified. 

\textbf{Route Origin Validation (ROV)} is the action performed by ASes to validate the route announcement based on existing ROA records. If a prefix has ROA records but the announcing AS is not in any of the ROAs, then such announcement is RPKI-invalid, and a ROV-enforcing AS should drop such invalid announcement. However, many ASes do not strictly enforce ROV. Some ASes may not perform any ROV at all, and some other ASes perform ROV but do not strictly enforce it (e.g., only lowering the priority of the invalid path instead of dropping it completely) \cite{Reuter_Bush_Cunha_Katz-Bassett_Schmidt_Wahlisch2018}. 
Unlike ROA records which can be directly retrieved from the public database, the ROV status of an AS is not trivial to measure from an outsider's perspective, and there exist several works in measuring the ROV status of all ASes \cite{Du_Testart_Fontugne_Akiwate_Snoeren_Claffy2022, Li_Lin_Ashiq_Aben_Fontugne_Phokeer_Chung2023, Reuter_Bush_Cunha_Katz-Bassett_Schmidt_Wahlisch2018, Hlavacek_Shulman_Vogel_Waidner2023}.

\textbf{Importance of RPKI.} 
While we focus on origin validation (ROV) in this paper, RPKI is an essential part and building block for path validation as well. 
For example, BGPSec~\cite{lepinski2017bgpsec} and Path-End validation~\cite{Cohen_Gilad_Herzberg_Schapira2015,Cohen_Gilad_Herzberg_Schapira2016} rely on the use of RPKI. If an AS does not yet have any ROA record in RPKI, then it cannot proceed with deploying any further path validation mechanisms.


\subsection{Motivation and Goal}

Motivated by the increasing deployment of RPKI, we aim to design an RPKI-based relay selection approach for Tor to (1) provide a stronger protection to route hijacks compared to Counter-RAPTOR \cite{Sun_Edmundson_Feamster_Chiang_Mittal2017}, and (2) 
%
decouple the traditional client location dependency from guard relay selection to avoid inadvertent leaking of client location information. Since our selection algorithm relies more on ROA coverage information of guard nodes rather than client locations, an adversary cannot learn sufficient information about client location even if she can observe the guard selection process.

\textbf{Threat model.} Our threat model is an AS-level adversary who is capable of manipulating announcements of targeted prefixes to perform BGP origin hijacks \cite{Sun_Edmundson_Vanbever_Li_Rexford_Chiang_Mittal2015}, i.e., the adversary's AS number appears as the last hop in the AS path of the announced prefix. 
Potential adversaries include malicious network operators or nation states who announce the prefix of target Tor relays and intercept traffic to the Tor relays.
Since Tor guard relays can observe Tor clients IP information, this makes them high value targets for BGP hijacking attacks.
We focus on BGP origin hijacking because it's found to be the most commonly observed type of route hijacking attack \cite{Cho_Fontugne_Cho_Dainotti_Gill2019, Qin_Li_Li_Wang2022}, and they are easy to perform compared to other more sophisticated attacks~\cite{birge2019sico,sun2021securing}.


\section{RPKI Coverage of Tor Relays}


While we observe a rapid increase in RPKI deployment on the  Internet in recent years \cite{NIST}, the RPKI coverage on the Tor network is unknown. 
Therefore, we first perform a measurement study using historical RPKI and Tor consensus data from Jan 2021 to May 2024, to quantify the trend of RPKI coverage of Tor relays overtime. The measurement result is critical in guiding the relay selection strategy. 


\subsection{Datasets}

We describe the datasets used in our measurement study.

\begin{itemize}
    \item \emph{Tor consensus}: 
    We obtained hourly consensuses from Jan 1, 2021, 00:00:00 to May 31, 2024, 23:00:00 from Tor Metrics \cite{Consensuses} and extracted relevant relay information, such as relay flags, IP addresses, and consensus weight. 
    \item \emph{Routing data}: 
    RouteViews~\cite{RouteViews} provides historical BGP routing updates. 
    We obtained hourly snapshots from Jan 1, 2021, 00:00:00 to May 31, 2024, 23:00:00. We extracted routing information pertaining to Tor relay IPs, i.e., the origin AS announcing a prefix that covers a Tor relay IP.
    \item \emph{Route Origin Authorisations (ROAs)}: We obtained ROA coverage information from RIPE RPKI archive \cite{ROA_Database} from Jan 2021 to May 2024 and extracted the data in the format of AS number, prefix, and max length.
    \item \emph{Route Origin Validation (ROV)}: Given the complexity of measuring ROV deployment, there does not exist one standard measurement. Instead, we obtained ROV data from several major sources, as described below:

    \begin{enumerate}
    \item \emph{ROV monitor}: Reuter et al. \cite{Reuter_Bush_Cunha_Katz-Bassett_Schmidt_Wahlisch2018} first measured ROV deployment in a controlled environment. By monitoring BGP announcement of carefully crafted ROAs, the authors were able to identify ASes that actively drop invalid announcement and categorized them as ROV deployed. However, the experiment is heavily reliant on the assumption that ASes must be connected to the PEERING testbed or using a route server. This assumption severely limits the number of ASes the authors can investigate and thus produced a fairly small list of ASes.
    
    \item \emph{MANRS}: Du et al. \cite{Du_Testart_Fontugne_Akiwate_Snoeren_Claffy2022} analyzed public routing behavior of MANRS (Mutually Agreed Norms for Routing Security) and non-MANRS network ASes and inferred ROV deployment, on the network scale, based on whether network ASes drop or propagate invalid routing information. We applied the same methodology to our network on the AS level. Using the data from the paper, we inferred AS-level ROV deployment based on whether an AS drops or propagates a BGP announcement that perfectly matches a valid ROA payload. 

    However, an ambiguous scenario arises when a BGP announcement contains prefixes that is not found in ROA payload. This does not necessarily mean the AS propagating those announcements is not ROV covered. To clarify such ambiguousness, we decided to investigate both cases separately. Case 1 will include the list of all ASes that propagate only BGP announcements that perfectly matches a valid ROA payload. Case 2 will include the list of all ASes that propagate BGP announcements that perfectly matches a valid ROA payload plus those unknown prefix announcements. Intuitively, case 2 will include more ASes than case 1 and we view case 1 as the lowerbound and case 2 as the upperbound.
    
    \item RoVista \cite{Li_Lin_Ashiq_Aben_Fontugne_Phokeer_Chung2023} is an ongoing measurement platform developed by researchers from Virginia Tech, IIJ, RIPE NCC, and MANRS to measure the current deployment rate status of ROV. It identifies ASes which are reachable under RPKI-invalid prefixes using IP-ID side-channel. RoVista assigns a percentage score to each AS indicating the progress of ROV deployment based on how much RPKI-invalid prefixes are being filtered out. Using its API, we checked all ASes in our dataset and based on its RoVista score, and produced a list of ROV-covered ASes.

    \item Another measurement of RPKI filtering done by Hlavacek et al. \cite{Hlavacek_Shulman_Vogel_Waidner2023} attempted to send bogus, RPKI-invalid prefix announcements and observed how they are handled by different ASes. They measure ROV deployment on ASes based on whether those handcrafted invalid announcement are dropped. We obtained their dataset and constructed a list of ROV-coverage ASes.
\end{enumerate}
\end{itemize}

\subsection{ROA and ROV Coverage for Tor}
\label{subsec:coverage}

We measure ROA and ROV coverage both on a relay level (i.e., percentage of relays) and on a bandwidth level (i.e., percentage of total relay bandwidth). We perform the measurement separately for all Tor relays and for guard relays only.  
For ROA coverage, we extracted all relays from the consensus data and separated them into two groups based on their IP: IPV4 and IPV6. Then, for each group, we compared their IP against the ROA dataset at the corresponding date to check if they have ROA coverage at that time. 
For ROV coverage, we compiled lists of ASes that have ROV deployment from the above-mentioned sources and cross-compared them with the list of ASes where Tor relays are located.

\begin{figure}[h]
  \centering
  \includegraphics[width=1\linewidth]{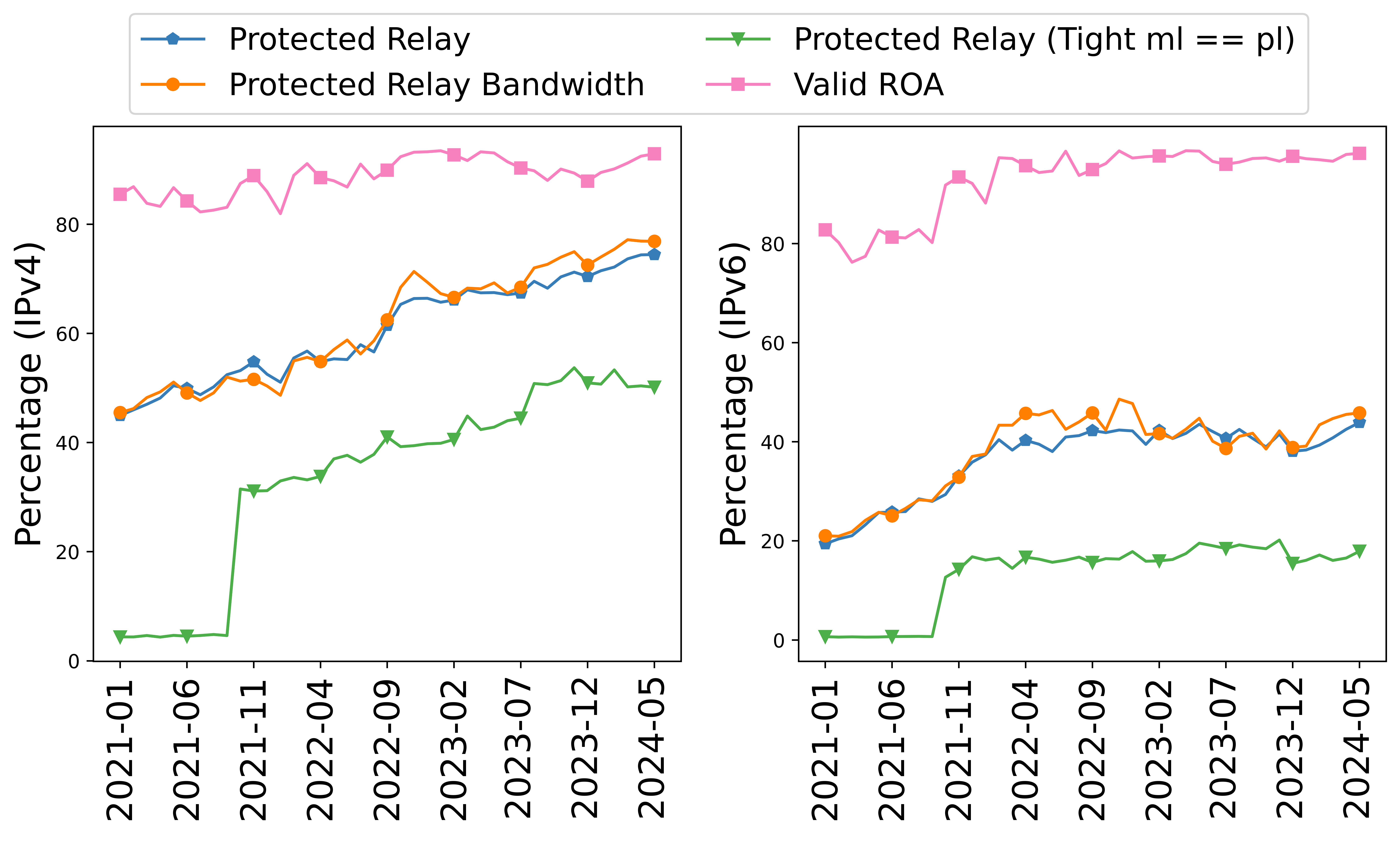}
  \caption{ROA Coverage and Validity for All Guard Relays}
  \vspace{-10pt}
  \label{ROA-Guard-All}
\end{figure}



\begin{table}
    \begin{tabularx}{\columnwidth}{|X|X|X|X}
        \hline
        Source                 & ROV-covered relays &  ROV-covered bandwidth \\
        \hline
        ROV Monitor & 2.696 & 2.530\\
        MANRS case 1 & 5.981 & 5.453\\
        RoVista & 36.675 & 33.517\\
        Hlavacek dataset & 52.008 & 49.093\\
        MANRS case 2 & 71.076 & 67.934 \\
        \hline
    \end{tabularx}
    \caption{Guard relay ROV coverage statistics (in \%)}
    \vspace{-20pt}
    \label{table: ROV_coverage}
\end{table}



\subsubsection{Coverage for All Tor Relays}

Figure \ref{ROA-Guard-All} shows the percentage of guard relays with ROA coverage from January 2021 to May 2024, using the consensus from the first hour of the first day of every month. The left figure shows percentage of ROA coverage for all IPv4 guard relays and the right figure displays the same information for IPv6 guard relays. 
In addition to showing ROA coverage, we also measure a stricter form of protection where max length equals the prefix length in the ROA record (green triangle). 
We also cross check with RouteViews data to validate whether the AS originating the prefix is valid based on the ROA record (pink square). 

\textbf{High IPv4 ROA coverage and low IPv6 ROA coverage.} Observe that close to half of guard relay IPv4 addresses were covered by ROAs in January 2021 (blue pentagons), and the percentage grows steadily throughout the years and reached 74.46\% in May 2024. On the other hand, the ROA coverage for guard relay IPv6 addresses is much lower. In January 2021, there were only 19.41\% of guard relay IPv6s that had a valid ROA. By the end of May 2024, the percentage has more than doubled, reaching 43.89\%, but still much lower compared to relay IPv4s.
We observe a similar trend in the bandwidth coverage for both IPv4 and IPv6 guard relays, and the percentage of guard relay bandwidth covered by ROA for IPv6 guard relays grew more than that for IPv4 guard relays. 


\textbf{Very high ROA valid announcement percentage.} 
The vast majority of route origins observed in RouteViews data for prefixes with ROAs are valid (pink squares). On average, more than 90 \% of prefixes covering IPv4 guard relays are valid. In May 2024, this percentage reaches approximately 92.91\%. This is also the case for IPv6. 
We then take a closer look at the reasons for invalid cases. Among all IPv4 announcements from all guard relays, 9.57\% of them have a mismatched ASN, 1.24\% have a prefix mismatch and 0.21\% have both a mismatched ASN and prefix. The distribution is similar for IPv6 guard relays. These mismatches may not necessarily be attacks, which could also be due to configuration errors.



\textbf{Few perfectly matched ROA payload.} The curve of green triangles shows the percentage of guard relays with a valid ROA with a tighter restriction: max length in ROA must equal the prefix length of the relay. This will force the relay announcement to be exactly the same as in the ROA database, which can provide an additional level of security against sub prefix BGP hijacking\cite{Sermpezis_Kotronis_Gigis_Dimitropoulos_Cicalese_King_Dainotti2018}. This percentage is much lower compared to without the strict restriction, which is as expected. Both IPv4 and IPv6 curves had a huge jump in Oct 2021. This is due to a lack of ROA coverage data for a significant portion of relay prefixes prior to October 2021.

\subsubsection{Coverage for All Relays}
Figure \ref{ROA-All} in Appendix \ref{ROA_coverage_all} displays the same information as Figure \ref{ROA-Guard-All}, but for all Tor relays. We observe the same trend for all relays for the period from Jan 2021 to May 2024.

\subsubsection{ROV Coverage}
Table \ref{table: ROV_coverage} summarizes the ROV coverage statistics for all guard relays using different sources of ROV  information. Due to the difference in sources, we can see the deployment rate varies, ranging from less than 3\% to over 70\%. Nevertheless, across all datasets, we can see that ROV bandwidth coverage is roughly at the same level with ROV relay coverage.




\textbf{Takeaway.} Our measurement study showcases a growing trend of ROA coverage for Tor relays, reaching 83.51\% of all relays in May 2024. This observation opens the possibility for clients to take advantage of the existing RPKI infrastructure by leveraging the ROA-protected guard relays to defend against route hijacking attacks. Additionally, while the current ROV deployment is much lower than ROA, it provides an opportunity to better leverage the existing ROV deployment to achieve a stronger protection against the route hijacks. 
With these insights, we explore the benefit of considering ROA coverage of relays in the guard relay selection process in Section~\ref{sec:discount}, and we take a step further to examine the potential of fully utilizing both ROA and ROV deployment statuses in the guard relay selection in Section~\ref{sec:matching}.

\section{Discount Selection Algorithm}
\label{sec:discount}

A natural way to defend against route hijacking attacks in Tor is to make use of the RPKI. 
It has the advantage of providing cryptographically-validated information on the validity of prefix/AS pairs (compared to the probabilistic inference in Counter-RAPTOR), and more importantly, moves away from client location-aware relay selection which leads to information leakage. 

Towards this goal, we propose a simple yet effective \emph{Discount Selection Algorithm} to increase the likelihood of choosing a ROA-covered guard relay. 
In a nutshell, the consensus weight of non-ROA-covered guard relays will be discounted. Choosing a ROA-covered guard relay provides the protection in the case of an attack, where the attack route will be dropped by ROV-enforcing ASes and consequently client traffic is still routed to the correct origin AS. 
Furthermore, such scheme may motivate large relay operators to host relays in ROA-covered prefixes. 

\subsection{Discount Factor}
In vanilla Tor, relays are selected with probability proportional to the consensus weight, which is determined by the relay bandwidth. 
The Discount Selection algorithm applies a \textbf{discount factor} $d (0 \leq d \leq 1)$ to the consensus weights for all guard relays. For relays with ROA coverage, the discount factor will be 1, i.e. the weight remains the same; for relays without ROA coverage, their consensus weight is multiplied by a factor of $d$, effectively reducing the probability that those relays will be chosen. When $d = 1$, the Discount Selection algorithm is exactly vanilla Tor, while when $d = 0$, all clients are forced to choose ROV-covered relays.

\textbf{Network load and available relay bandwidth.}
The Discount Selection algorithm works well when the Tor network is not fully saturated. If all relay bandwidth is fully utilized, then there is no room for shifting traffic to ROA-covered relays without causing excessive performance degradation. In other words, the appropriate value for discount factor is affected by the network load in regards to total relay bandwidth. In the fully saturated scenario, a discount factor of 1 should be applied. 

\textbf{Distribution of discount factor to clients.}
Tor directory authorities receive reports of available bandwidth and current capacity from relays, which are at a natural position to determine the discount factor based on such information, combined with the ROA status of the relays. 
Therefore, it is a natural choice for the directory authorities to update and distribute the discount factor in the hourly consensus file, e.g., as an additional parameter in the consensus which will be used by the Tor client code when computing final weights for guard relay selection. 


\subsection{Security Evaluation}
\label{subsec:discount_security}

To measure the effectiveness of the Discount Selection algorithm, we implement a local python simulation of 1 million  randomly-generated clients performing guard relay selection using the Discount Selection algorithm. Since the discount selection algorithm is client-agnostic, i.e. it does not depend on inherent client characteristic such as IP address or geographic location, we randomly sampled 1 million clients with random IP addresses.

\subsubsection{Relay Selection}
In each simulation run, the client first loads in the consensus and checks against the ROA database to identify ROA-covered guard relays. Then, the non-ROA-covered relay weight is adjusted based on the discount factor. Finally, the probabilistic relay selection is performed similar to vanilla Tor. 

%

\subsubsection{Load Balancing}
Tor performs load balancing by actively measuring available relay bandwidth, which is then combined with relay's self-reported bandwidth to compute the consensus weight. 
In a situation where the relay's bandwidth is fully saturated by user traffic, the measured bandwidth will be negatively affected, which will then result in a lower consensus weight until its traffic load decreases. 
Faithfully mimicking such practice will involve simulation of actual user traffic, which is a task achieved by the Shadow simulator(we will use it in Section~\ref{subsec:shadow}). 


To take into consideration of such load balancing in our lightweight simulation for security evaluation, we implement a \emph{dynamic load balancing} that aims to ensure that the clients can utilize sufficient bandwidth in their selected relays. This is especially important when the network load is high and relays are nearly saturated, any shift in relay selection due to the discounted algorithm may result in a client not getting enough bandwidth. 

More specifically, we use the the \emph{average bandwidth per relay per client} as the proxy to learn whether a relay may be saturated and not have sufficient available bandwidth for additional clients. 
During client relay selection, the simulation counts the number of clients in the network and computes the average bandwidth per relay per client. This number is updated as more clients join the network and select their relay. 
If a given client's share of bandwidth from its selected relay (i.e., relay bandwidth divided number of clients who have selected this relay) falls below the overall average, the client will need to perform relay selection again and select another relay. 
The selected, overloaded relay will then reduce its weights by a factor of the chosen relay's per client bandwidth over the network average client bandwidth. This factor takes into consideration of the current load per relay and is used to balance each relay's selection probability based on the number of current client connections, assuming an even share of bandwidth. As more and more relays report network overload, eventually all of them will reduce their weights and but from a network's perspective, the weight distribution will stay largely the same as the original. All weights are scaled down but stay proportionally the same. This enables the simulation to handle 1 million clients performing relay selection without worrying about "running out of bandwidth".



\textbf{Load Factor.}
As mentioned above, the network load is an important factor in determining the appropriate discount factor and in simulating the load balancing across relays. 
Thus, we introduce a parameter named load factor $l (0 \leq l \leq 1)$ into the simulation. This parameter indicates how much total guard relay bandwidth is utilized. For example, $l = 0.8$ means the guard relay bandwidth is currently 80\% utilized and have up to 20\% of its total bandwidth available. This parameter is between 0 and 1, with 1 indicating the network is currently fully saturated. 
In such case, any shift from the bandwidth-based vanilla Tor relay selection may result in performance degradation. 
Based on Tor metrics~\cite{Tor_Metrics_BW}, the actual load in Tor guard relays (i.e., fraction of consumed bandwidth over advertised total bandwidth) is around 0.45. 
In our simulations, we show results using load factors ranging from 0.3 to 1. 

\subsection{Results and Analysis }
\label{subsec:discount_result}

\subsubsection{Experiments and Metrics}
We run simulations of 1 million clients performing relay selection using Tor network data from January 01, 2021 to May 31, 2024. To address the inherent randomness, each simulation is run 100 times and final results are aggregated over the average. Because the consensus is updated on an hourly basis, for each month, we used the consensus of the first hour of the first day for simplicity. 
To evaluate the effectiveness of the discount selection algorithm, we use the \emph{percentage of clients that choose a relay with ROA coverage} as the evaluation metric. 

\subsubsection{Client ROA Coverage Rate}
\begin{figure}[htp]
  \centering
  \includegraphics[scale=0.5]{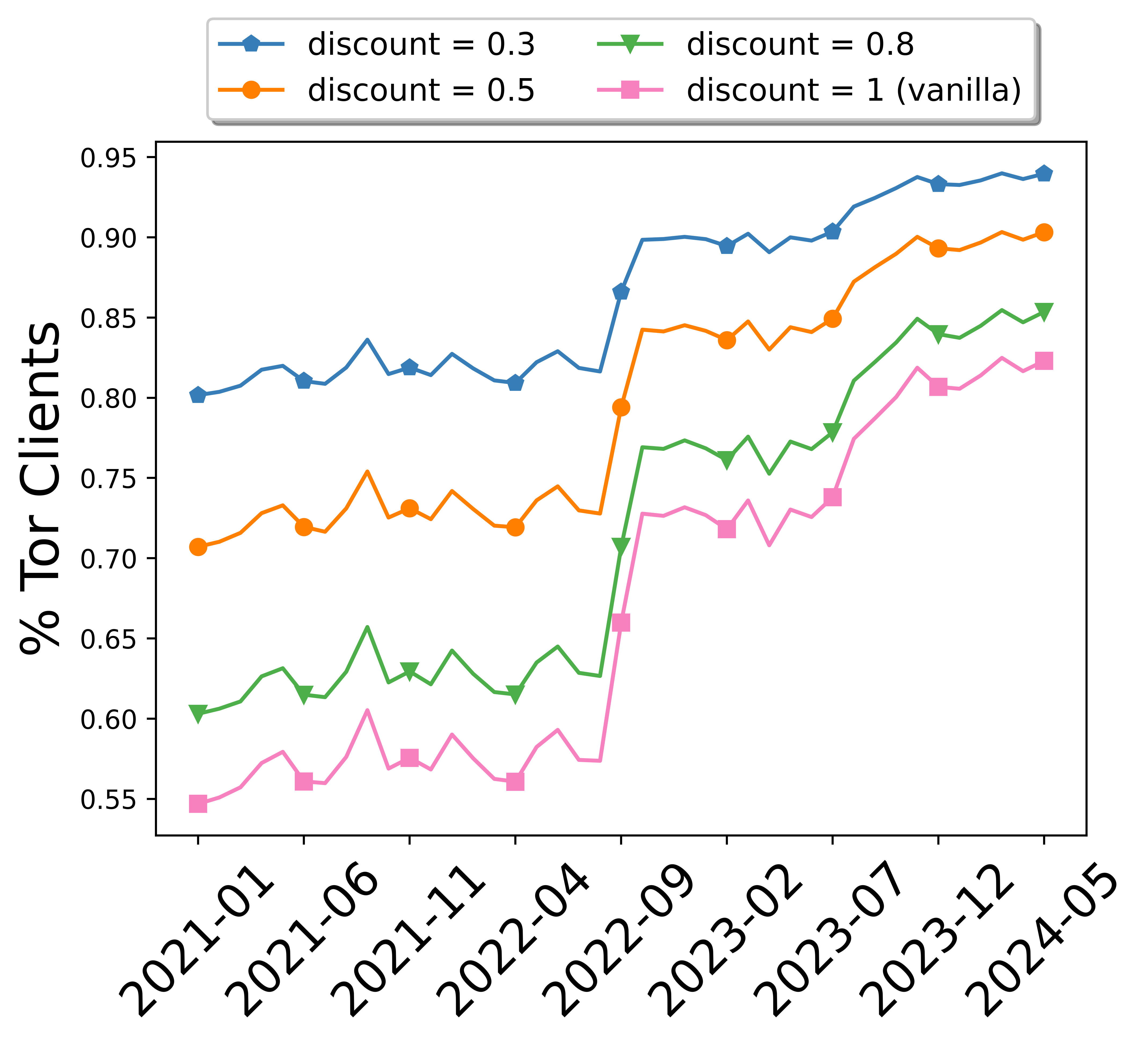}
  \caption{Percentage of clients with ROA covered guard at different discount value over time}
  \vspace{-8pt}
  \label{Discount-ROA}
\end{figure}

Figure \ref{Discount-ROA} shows the percentage of clients with ROA-covered guard with different discount factors when applying the Discount Selection algorithm. We used a discount factor of 0.3 (blue pentagon), 0.5 (orange circle) and 0.8 (green triangle) to represent high, medium and low discounts, respectively. We also included vanilla guard selection as a baseline (pink square). Intuitively, the higher the discount (i.e., lower discount factor), the better the ROA coverage rate will be, as most clients will be forced to pick relays with ROA coverage. We can see that with a discount factor of 0.3 (blue pentagon line), the simulation is able to achieve more than 80\% ROA coverage at all times (reaching above 90\% after November 2022) compared to less than 75\% in vanilla Tor. 

We also notice a steep increase in the coverage around August 2022. 
Upon further inspection, an increase of around 80,000 IP prefixes with ROA coverage is found. This result is also corroborated by our relay ROA coverage measurement in Section~\ref{subsec:coverage}, where ROA coverage increases from 66\% in May 2022 to 74.5\% in October 2022. 

\subsubsection{Network Load and Discount Factor}
It is important to point out there does not exist a "best" value for discount factor. It depends on the subjective goal of load balancing and achieving a balanced overall network load. 
In reality, the Tor directory authorities will learn the network load reported by relays and measurements, and need to determine the discount factor based on the current load. 
Next, we show the effect of various discount factors on the bandwidth utilization. 
Figure \ref{Discount-Load} in Appendix \ref{Load_x_discount} shows the expected bandwidth utilization under various combinations of initial load factor and discount factor using the consensus from 05/01/2024 00:00:00. 


\textbf{When initial load is low, it becomes the limiting factor.} A flat line indicates the current network has not achieved its full load capacity yet. We can see when initial load is lower than 0.8, the actual load utilization is equal to the initial load factor no matter the discount applied.  This is intuitive since if there is little traffic then it matters little which discount we applied to relays without ROA as the relays with ROA are capable of handling all incoming traffic. This is further validated by Figure \ref{ROA-Guard-All}, where we can see almost 80\% of IPv4 relay bandwidth is covered by ROA on 05/01/2024.

\textbf{When initial load is high, both discount and load factor together dictate the network maximum expected utilized bandwidth.} When initial load is much larger, both discount factor and load factor play a role in the expected bandwidth utilization. For lines with high initial load ($l > 0.8$), we can see there exists a "kink" in the line in Figure \ref{Discount-Load}. This kink is the effective maximum throughput and is at the level of the discount factor. This is the best-case scenario where applying discount has no impact on the total throughput as vanilla Tor and beyond this point, further discount will start to reduce network throughput. We can see for when initial load is at 0.9, the line has a slope when discount is low and turns flat when discount is at or above 0.4. This suggests the network is capable of handling all incoming traffic when discount applied is at or above 0.4. This is consistent with our expectation that the stronger the discount (lower discount factor), the lower the network throughput.

Based on the above observations, to fully utilize bandwidth under a discounted network, it is important to find a discount factor by considering the network load and discount factor together.
For our next simulation, we choose the smallest discount factor (i.e., largest discount) that can still fully utilized bandwidth (i.e., the "kink" in the line in Figure~\ref{Discount-Load}). For simplicity, we will refer to it as the "optimal discount".  
We show how such discount factor changes with different load factors and different Tor consensuses overtime.

\begin{figure}[htp]
  \centering
  \includegraphics[scale=0.5]{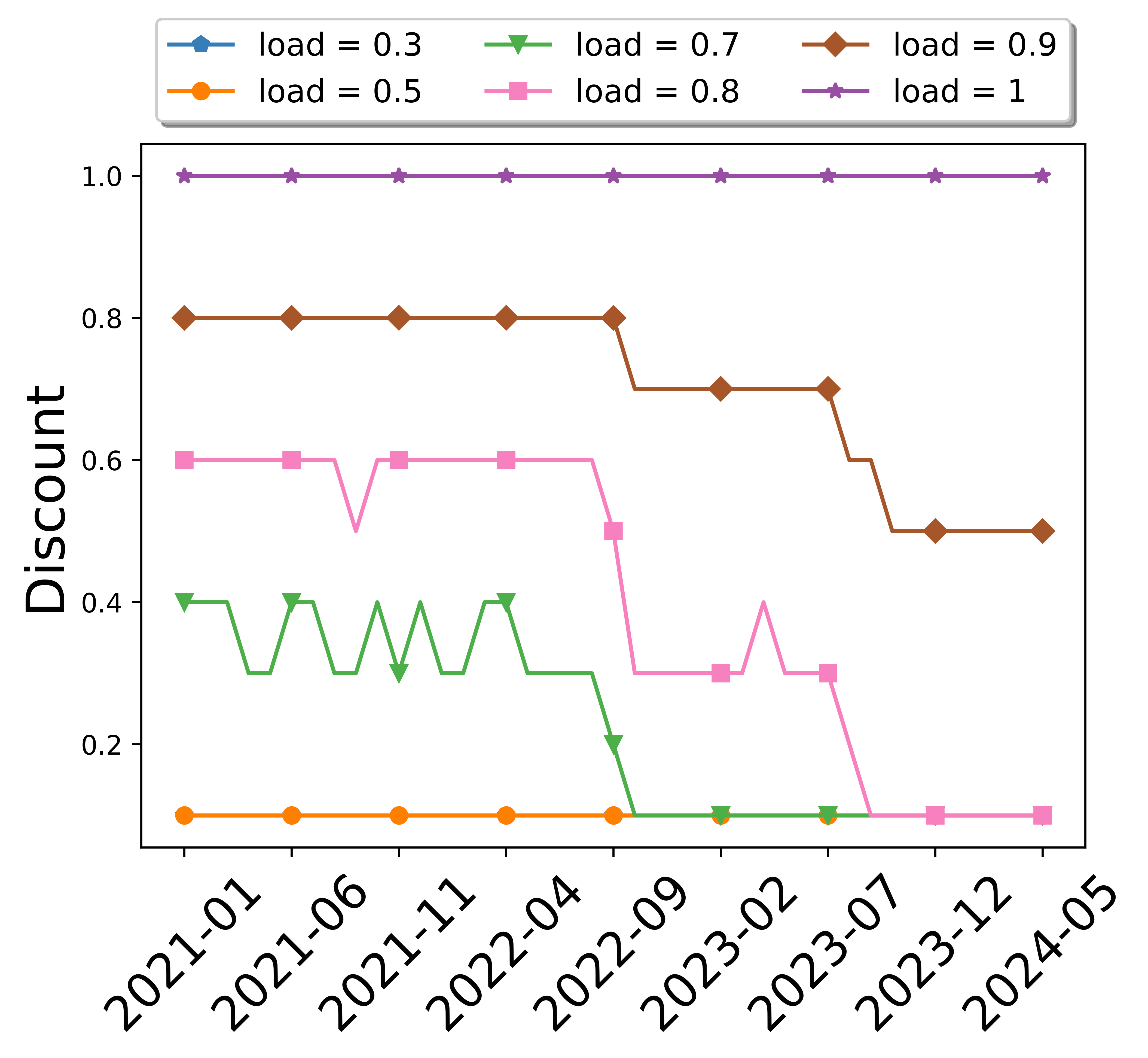}
  \caption{Discount factors with various load factors over time}
  \vspace{-8pt}
  \label{Discount-Optimal}
\end{figure}

\textbf{Variation of discount factors.} Figure \ref{Discount-Optimal} shows the "optimal" discount factor under various load factors through the three-year period. We can see that when load factor is low enough (less than 0.5, which is the case in real Tor network), we can adopt a very small discount factor since we have enough bandwidth for all traffic to select relays with ROA coverage. This can be verified from Figure \ref{ROA-All} in Appendix \ref{ROA_coverage_all}, which shows the percentage of bandwidth covered by ROA for both IPv4 and IPv6. When load factor is low, the percentage of bandwidth covered by ROA alone is enough to cover all incoming traffic.
On the other hand, when bandwidth is 100\% utilized, no discount is possible without overloading the relays.
Furthermore, the discount factor for the same load factor drops around August 2022, due to the increase in relay ROA coverage.

The Shadow simulation of the Discount Selection algorithm results are included in Section 5.4.


\textbf{Takeaway.}
The Discount Selection algorithm is a simple yet effective approach that can be performed by directory authorities to enable more clients to use ROA-covered relays with very minimal changes to Tor. 
Nevertheless, it is important to keep in mind that there does not exist a "best" discount factor. It is affected by the number of ROA-covered relays and the current network load. 
We show simulation results as a guidance on understanding the effects on discount factor, but discretion is needed to achieve the ideal outcome when used in a real-world scenario.

\section{Matching Selection Algorithm}
\label{sec:matching}

The Discount Selection algorithm in Section~\ref{sec:discount} 
provides a lightweight approach by only considering the ROA coverage of guard relays. 
However, it's worth pointing out that to effectively prevent BGP-hijacking attacks, it is also important that ASes perform ROV to drop invalid route announcements. 


Therefore, to fully utilize the security benefits provided by ROA and ROV, we propose a \emph{Matching Selection Algorithm} that considers both ROA coverage and ROV deployment status. 
In a nutshell, the algorithm aims to match pairs of ROA-covered ASes (client or guard) with ROV-covered ASes (client or guard). Such pairs can fully utilize ROA through proper ROV and thus achieve a stronger level of security against BGP-hijacking attacks. 


\subsection{Methodology}

\subsubsection{Goal and Overview}


Our primary goal is to match ROA-covered clients with guard relays in ROV-enforcing ASes, and vice versa. However, the number of ASes that have ROV deployed is much lower compared to the number of ASes with ROA coverage. Therefore, it is unrealistic to achieve ROA-ROV matching on all client-relay pairs, especially when network load is high. Additionally, forcing all traffic through ROV-enforcing relays may introduce additional vulnerabilities such as guard placement attacks because the ROV-enforcing relays will have significantly higher probability of being chosen, 
which will enable a malicious relay to obtain more than its fair share of information. 

\textbf{Limitation of Discount Selection algorithm.} An intuitive idea is to improve upon our discount selection algorithm and introduce a second discount factor for non-ROV-enforcing relay. Using the same methodology, we can discount different categories of ASes each with a unique combined discount factor, e.g. an AS with ROA deployed only will be discounted by $d_{1}$, an AS enforcing ROV only will be discounted by $d_{2}$, and an AS with neither ROA nor ROV will be discounted by $d_{1} \cdot d_{2}$. Recall from Section~\ref{sec:discount} that the discount factor is affected by factors such as the network load and relay ROA coverage, it is significantly more complex here when considering two discount factors and considering ROA/ROV for both clients and relays. 
Therefore, we need to devise a more robust way to determine weights in relay selection to achieve the goal of ROA-ROV matching of clients and relays while considering network load.

\textbf{CLAPS framework.} CLAPS \cite{Rochet_Wails_Johnson_Mittal_Pereira2020} is the state of the art on location-aware Tor path selection. In CLAPS, an optimization process acts as the directory authorities, recalculates relay weights to satisfy their objective and then distributes the new weights to clients from specific groups. We adopted the same principle in developing our matching algorithm with two key differences: (1) Our algorithm does not require client location and thus is not location-aware; (2) Our algorithm focuses on guard relay selection and does not alter the selection algorithm for middle relays and exit relays. 


\textbf{Overview of Matching algorithm.} The Matching algorithm works as follows: the algorithm first categorizes all clients into four subcategories, with each subcategory denoting a combination of the client's ROA and ROV deployment status: ROA-deployed only, ROV-deployed only, both ROA and ROV deployed and neither deployed. For each category, the algorithm recomputes a set of weights for all relays in the network. When clients connect to the network, the algorithm performs ROA and ROV checking to identify which subcategory the client belongs to, and then sends the set of optimized weights for the corresponding subcategory. Then the client proceeds to the relay selection process as usual.


\subsubsection{Weight Computation}
Based on our objective, we will frame the matching problem into a linear optimization problem. First, we will define a set of reward functions for matching a pair of client and relay ASes based on their ROA/ROV status. Then our goal is to maximize the total reward by performing all clients relay selection and adding up the individual pair-wise reward. A ROA-ROV matched pair increases the objective value while a non-matched pair reduces the objective value by a penalty, which is further differentiated based on whether ROA or ROV coverage is missing. 

To express our objectives more accurately, we define the parameters and additional variables used here: 

\begin{enumerate}
    \item Let $0 \leq l \leq 1 $ be the load factor of the current network, representing the percentage of total bandwidth of the network that is currently utilized. This value is equivalent to the load factor in Section~\ref{sec:discount}, which indicates the current network load. We capped $l$ at 1 because the assumption of the optimization is that there is still "unused" bandwidth and there is room for improvement for currently under-utilized network. If this parameter goes above 1, it will force the optimization to reduce all bandwidth for all relays, which will end up producing a worse performance than what we begin with. This is undesirable and thus the cases where $l$ goes above 1, such algorithm should not be used. Note that the current guard relay load is around 0.45.
    \item Let $\theta \geq 1$ be the maximum factor by which any guard relay’s selection probability can increase over vanilla Tor. This factor limits the susceptibility to a relay placement attack and is commonly set to 5 \cite{Wan_Johnson_Wails_Wagh_Mittal2019}.
    \item Let $\mathcal{S} = \{roa, rov, both, neither \}$ be the set of all subcategories an AS belongs to. Each AS can only belong to one subcategory at any given time. 
    \item Let $\mathcal{R}$ be the set of all guard relays and let $\mathcal{R}_{s}$ be the set of guard relays in subcategory $s$, e.g. $\mathcal{R}_{roa}$ represents all the guard relays that have ROA coverage only.
    \item Let $P(r_{s_{r}},c_{s_{c}})$ be the penalty/reward function mapping the combination of a relay $r$ with status $s_{r}$ and a client $c$ with status $s_{c}$. This penalty/reward is directly applied after a client $c$ has chosen relay $r$ and the client-relay pair has been formed. To better illustrate the distribution, we list all possible 16 combinations below:
    \begin{enumerate}
        \item The pair is perfectly matched both ways. Both client and relay are both ROA-covered and ROV-covered.
        \item The pair is matched but not both ways. This includes the following cases:
            \begin{enumerate}
                \item Client has both ROA and ROV coverage but relay is only ROA-covered and vice versa.
                \item Client has both ROA and ROV coverage but relay is only ROV-covered and vice versa.
                \item Client has ROA coverage only and relay has ROV coverage only and vice versa.
            \end{enumerate}
        \item The pair is not matched. This includes the following cases:
            \begin{enumerate}
                \item Client has ROA coverage only and relay has ROA coverage only. Vice versa.
                \item Client has ROA coverage only and relay has neither ROA nor ROV coverage. Vice versa.
                \item Client has ROV coverage only and relay has ROV coverage only. Vice versa.
                \item Client has ROV coverage only and relay has neither ROA nor ROV coverage. Vice versa.
                \item Client has neither ROA nor ROV coverage and relay has neither ROA nor relay coverage either.
            \end{enumerate}
    \end{enumerate}
    
    \textbf{Reward/Penalty structure.} To quantify the reward/penalty, we use two discount factors for missing ROA coverage and missing ROV coverage, respectively. For an AS missing both ROA and ROV coverage, we simply multiply the two discount factors to obtain a smaller discount value, i.e. more penalty. Then to find the reward/penalty for a client-relay pair, we simply multiple each AS's reward/penalty together to find the final reward/penalty. Note here, the discount value for missing ROA is smaller than that of missing ROV, i.e. missing ROA is penalized more than missing ROV, since we value missing ROA coverage is more serious than missing ROV coverage. 
    
    Since our objective is to find as many ROA-ROV matched client relay pairs as possible, this dictates that the reward for a matched pair should always be higher than that of a non matched pair, no matter which case they each belong to. However, using only two discount factors may produce edge cases. Consider the two following pairs: (a) a ROA and ROV covered client with a non-ROA, non-ROV relay; (b) a ROA-only client with a ROV-only relay. Pair (b) is clearly more desirable than pair (a) since we successfully matched pair (b) but reward-wise, both pairs share the same penalty. Therefore, to maintain consistency across pairs while achieving the original goal, we introduce a third variable called \textbf{matching bonus}. The bonus is applied to a pair whenever the pair is matched, no matter if it is a perfect match. This will make the reward structure strictly decreasing in the order of a perfect match, a match with missing ROV, a match with missing ROA, a non match missing ROV, a non match missing ROA and finally, a non match missing both. 
    
    It is worth pointing out that either a positive (reward) or negative (penalty) function works here. If the function is all negative, the objective becomes finding the minimal penalty instead of maximal reward. In the rest of this section, we refer to the function as the reward function.

    \item Let $w_{r, s}$ denote the weight for \textbf{relay} $r$ when paired with a \textbf{client} with status $s$. It is important to point out here that $s$ in the notation refers to the client status, \textbf{NOT} the relay status and $w_{r,s}$ will be the variable the linear program optimizes. 
    
    \item Let $b_{r}$ denote the bandwidth for relay $r$. The bandwidth is the original value recorded in the consensus file and is fixed throughout the optimization. 

    \item Let $T_{s}$ be the percentage of clients in the Tor network that have status $s$, e.g. $T_{roa}$ represents the percentage of clients in the Tor network that have ROA coverage only. The value of $T$'s is also fixed throughout the optimization. 
\end{enumerate}

\textbf{All weights are normalized.} To better formalize our goals, we express all original weights and optimized weights in normalized form. A byproduct of such normalization is that we can use the normalized weight directly as the probability of a relay being chosen. Another way to interpret our objective is to maximize the weighted average of reward function over all possible relay and client ROA/ROV statuses. This means the weight not only depends on the normalized probability but also the client distribution.

\subsubsection{Objective Function}
Now we can express all our goals in a linear program. Our primary objective is to maximize the cumulative reward from client relay selection. Our secondary goals can be expressed as constraints in the linear program: (a) We want to maintain a load-balanced network. This constraint is two-fold. Individually, no single relay should take on more traffic than its maximum bandwidth allowed; globally, the entire network should not take on more traffic than its bandwidth. (b) The network should remain $\theta$-GP secure. No single relay should have more than $\theta$ times advantage of being selected compared to that in vanilla Tor.

\begin{equation} 
\label{eqn:1}
\max_{\bm{}} \quad
\displaystyle\sum\limits_{s \in \mathcal{S}} \sum\limits_{r \in \mathcal{R}} w_{r, s} P(s_{r}, s_{c})
\end{equation}
\begin{subequations}\label{eq:subeqns}
\begin{align}
\text{subject to}\quad
&\forall \hat{s} \in \mathcal{S} \displaystyle\sum\limits_{r \in \mathcal{R}} w_{r, \hat{s}} =  1 \label{eq:subeq1} \\
&\forall \hat{r} \in \mathcal{R} \displaystyle\sum\limits_{s \in \mathcal{S}} T_{s} w_{\hat{r}, s} \leq \frac{b_{\hat{r}}}{l} \label{eq:subeq2}\\
& \forall \hat{r} \in \mathcal{R} \displaystyle \frac{w_{\hat{r}, s}^{}}{ \sum\limits_{r \in \mathcal{R}} w_{r}} \leq \frac{\theta w_{r}^{'}}{\sum\limits_{r \in \mathcal{R}} w_{r}^{'}} \label{eq:subeq3}
\end{align} 
\end{subequations}



Our linear program (\ref{eqn:1}) expresses our main objective. We sum over all the rewards for all relays for all combinations of subcategories and its matched client subcategories. Equation (\ref{eq:subeq1}) is the global load constraint. We make sure the sum of all weights for all relays within each subcategory should not exceed network load. We want to fully utilize all relays under every subcategory without overloading the global network. Since we are using normalized weights, the network load is simply 1. Equation (\ref{eq:subeq2}) bounds the individual relay weight by mandating that its weights, summed over all four subcategories, should not exceed its maximum allowable weight, which is calculated by its bandwidth divided by the load factor. Here we need to use the client distribution as the probability instead of relay distribution. This means that the overall weighted average is a weighted average of relay weights averaged on client coverage distribution. Finally, equation (\ref{eq:subeq3}) enforces the guard placement constraint that no single relay should have more than $\theta$ times the probability of being chosen compared to vanilla Tor. This is equivalent to limiting the optimization (weighted) weight to not exceed $\theta$ times the normalized weight, for all relays.

\textbf{One optimization for all four subcategories.} One crucial question is to whether run one optimization using all relays for all clients or run four optimizations, each using all relays for only a subcategory of clients. Since the same relay is valued differently by different clients based on their own ROA and ROV status, the weights distribution will be different from subgroup to subgroup. Intuitively, four optimizations is reasonable. Nevertheless, given the constraint (2a) and (2b), each relay's weight under one subcategory of clients is dependent on its weight under the other subcategories of clients as they share a global cap. Therefore, it will be inappropriate to run separate optimizations despite the potential runtime speedup. Instead, the optimization takes four times number of relays as input variables, with each quarter representing the set of relays' weights under a subcategory of clients and the optimization optimizes all relays weights simultaneously for all subcategories.

\subsection{Client Generation and Parameters}

\subsubsection{Difference in Client Distribution between Discount Selection Algorithm and Matching Selection Algorithm} A crucial difference between the Discount Selection algorithm and the Matching Selection algorithm is whether clients are status-agnostic. In Discount algorithm, no matter which ROA and ROV status a client has, it performs the same relay selection process. Therefore, it suffices to randomly assign IP addresses to clients for the client generation step. Nevertheless, the same process cannot be applied when simulating Matching algorithm simply because randomly assigning client IP addresses may produce a sample ROA/ROV status distribution that is not representative of the global ROA and ROV distribution of the real Tor clients and can skew our results.

\subsubsection{Client Generation}
To generate a list of clients that mimic the distribution of real Tor clients, both in geography and ROA/ROV status, we obtain Tor user distribution by country from Tor Metrics~\cite{Tor_Metrics_Users} and for each country, identify all ASes that are geographically located in it by extracting the Inferred AS to Organization Mapping Dataset from CAIDA~\cite{AS2Org}. Then, for each AS, we obtain all prefixes it controls using the RouteViews data. Finally, we check each prefix against the ROA and ROV database and tally up the number of IP addresses under each prefix for all ASes in a country to obtain that country's ROA and ROV status distribution. Combined with the geographic distribution, we now have a global ROA and ROV distribution by country.

To generate a client, we first randomly assign its country using the Tor user distribution by country data. Then within that country, we randomly assign its ROA and ROV status using that country's ROA and ROV distribution. Since the Matching Selection algorithm only requires a client's ROA and ROV status, this enables us to completely bypass the client AS and IP generation and provides a significant speedup to the overall simulation.

\subsubsection{Parameters}
In this subsection, we discuss our choice of parameters used in the simulations.
\begin{enumerate}
    \item $\theta$, the maximum factor by which any guard relay’s selection probability can increase over vanilla Tor, is set to 5, following common practice \cite{Wan_Johnson_Wails_Wagh_Mittal2019} to reduce the susceptibility to relay placement attacks.
    \item $l$, the initial load factor, is set to 0.8. This parameter is chosen statistically. As discussed in Section 4, the discount factor only comes into effect when there is high enough initial traffic volume. The same principle applies to the reward structure. Setting this parameter to 0.8 provides us: (a) initial high traffic volume so our reward structure can influence the relay selection process; (b) a more dynamic simulation as our dynamic load balancing will be active more frequently and client churn will be more impactful. It can be changed to other values to adopt to custom network loads.
    \item The penalty/reward structure, denoted as $d_{1}$, $d_{2}$ and $B$, representing the discount factor for missing ROV, discount factor for missing ROA and the matching bonus respectively. From the discussion in section 5.1.2, we know that (a) the penalty for missing ROA is more severe than missing ROV, i.e. $0 \leq d_{2} < d_{1} \leq 1$; (b) the matching bonus must be positive, i.e. $B > 1$. Additionally, the strictly decreasing reward structure dictates that $d_{1} \cdot d_{1} < B \cdot d_{2} < B \cdot d_{1} < B$. Solving for $d_{2}$ gives $\frac{d_{1} \cdot d_{1}}{B} < d_{2} < d_{1}$. This inequality puts a further lowerbound on $d_{2}$ in terms of $d_{1}$.

    We experimented on various combinations of $d_{1}$, $d_{2}$ and $B$ with varying load factors and examined the matched rate under each specific parameters combo, and the matched rate improvement compared to vanilla Tor. All simulations were on the same set of 1 million clients generated using the method described in section 5.2.2. The detailed results are in Appendix \ref{Appendix_A}.

    From the results, we can observe that with a given load factor, the impact of different reward structure on matched rate is limited, percentage-wise. However, given the 1 million client base, a 0.0001 difference still translates to 1000 more matched client-relay pairs. Overall, we find a combination of 0.9-0.7-1.5 reward structure provides the best matched rate as well as improvement in matched rate, under various load factors.

    For the rest of the paper, unless otherwise specified, we used the following set of parameters: $l = 0.8$, $\theta = 5$, $d_{1} = 0.9$, $d_{2} = 0.7$ and $B = 1.5$.
\end{enumerate}


\subsection{Security Evaluation}
To measure the effectiveness of the Matching Selection algorithm,
we perform simulation of 1 million randomly-generated
clients performing guard relay selection, using the method described in section 5.2.2. 
For each case, 100 runs of 1 million clients selecting guard relays is performed using our Matching algorithm and vanilla Tor guard relay selection. The same set of clients and the same consensus from 2024-05-01 00:00:00 is used for all simulation runs.




\textbf{Multiple sources of ROV deployment status.}
As we have shown in Table~\ref{table: ROV_coverage} in Section~\ref{subsec:coverage}, the numbers of ROV deployment varies significantly from different sources. 
The ROV deployment monitor shows the lowest ROV deployment rate (<3\%), which was last updated several years ago. 
On the other hand, MANRS case 2 shows the highest ROV deployment rate (close to 70\%) due to the inclusion of all unknown cases, which is meant to serve as an upper bound of ROV-enforcing ASes. 



In our simulation, we consider all these five sources of ROV deployment status and show how the percentage of ROA-ROV matched pairs varies based on the ROV deployment status.

\subsubsection{Results}
\begin{figure}[h]
  \centering
  \includegraphics[width=\linewidth]{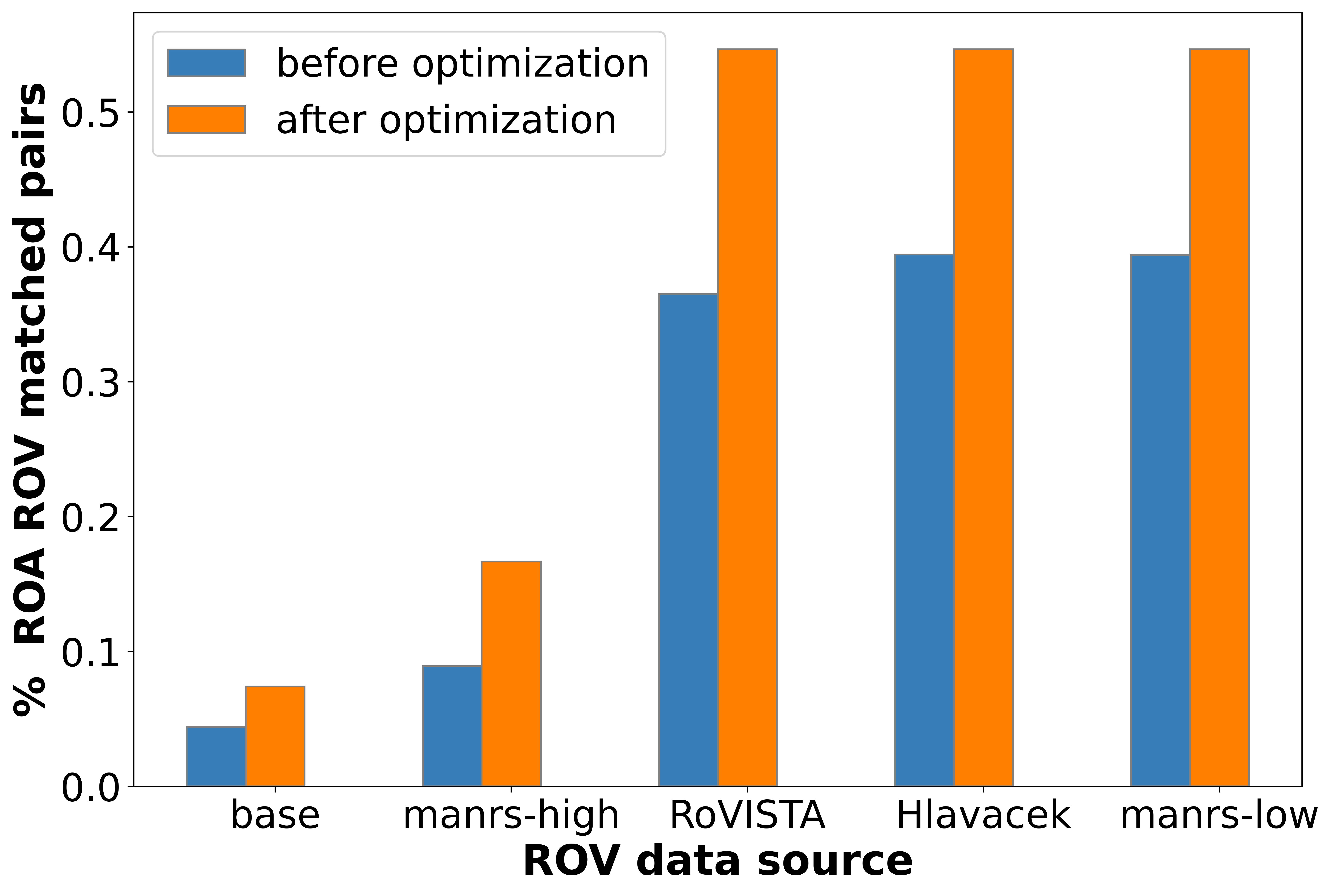}
  \caption{Percent of relay-client pairs with matched ROA \& ROV}
  \vspace{-8pt}
  \label{Matching-ROA-ROV}
\end{figure}

Figure \ref{Matching-ROA-ROV} shows the comparison in ROA-ROV match rates between vanilla Tor (blue bar on the left, before optimization) and matching algorithm (orange bar on the right, post optimization). 

Overall, the optimization is highly efficient, significantly boosting matched rates in all cases. Nevertheless, the degree of the improvement varies from case to case. The cases where there are lower pre-optimization rates leave more room for improvement and tend to benefit more compared to cases where there is already a high initial match rate. Furthermore, the match rate is heavily affected by the number of ROV-deployed ASes, especially when the ratio is low. 
This is evident from the rapid increase from base (ROV monitor) to RoVISTA. 
However, while Hlavacek and MANRS-case-2 include more ASes as ROV-deployed ASes, the increase in match rate is not significant. 
This is potentially due to the limited Tor bandwidth in those additional ASes, which does not provide a significantly higher amount of additional ROV-deployed bandwidth to match. 

\textbf{Takeaway.} The matching algorithm results in multifold increase in the percentage of ROA-ROV matching between clients and relays, which provides a stronger protection against route hijacks compared to the discount algorithm (ROA only). 

\subsection{Performance Evaluation}
\label{subsec:shadow}

To further measure the scalability and efficiency of our Matching algorithm, we implement the algorithm into Tor code and run a large scale simulation using the Shadow network simulator~\cite{Jansen_Hopper2011}. 
Shadow is widely used in the network research community and is highly regarded as a scientific way for Tor network traffic analysis due to its faithful simulation of various network conditions. All our Shadow simulations are performed on a 16-core virtual machine with 450 GB of ram and 500 GB of disk, running Ubuntu 22.04 on the FABRIC testbed~\cite{fabric}. All simulations are run with the baseline parameter described in section 5.2.3. Each configuration is run 10 times to account for randomness in the simulation~\cite{jansen2021once}. All simulations is running Tor 0.4.8.9. For simulation purposes, the modified Tor client will load in the new weights computed by the optimization based on its ROA and ROV status. In a real world scenario, there will be no changes required from clients as all the entire process is computed internally at the directory authorities. See detail discussion in section 6.3.

\textbf{ROV deployment status.}
For Shadow simulation, we choose to use the RoVista ROV database given its completeness of the measurement, which shows 
37\% of relays and 33.5\% of relay bandwidth is ROV-covered according to Table ~\ref{table: ROV_coverage} . 
Furthermore, the incremental benefit of match rate is limited even when using the other two larger ROV lists compared to RoVISTA. 


\subsubsection{Simulation Setup}
Following the model from \cite{Janse_Traudt_Hopper2018}, we staged all relay information using consensuses, server descriptors from May 2024. We also used Tor user-country data in the staging process. For simulation, we generated 10\% scale of the Tor network, involving 719 relays and 75234 Tor clients. Clients are generated using Tgen processes as detailed in \cite{Janse_Traudt_Hopper2018}. Simulation is performed as data transfer between Tgen instances such as simulation of a client transferring data from/to a relay. We have expanded the Tgen instances to run various data transfer sizes which include: 1MB, 5MB, 10MB, 50MB and 100MB, with the larger file sizes transfer having less probability, consistent with the probability observations from \cite{tmodel-ccs2018}.
Note that this scale is magnitudes larger than the scale used in CLAPS~\cite{Rochet_Wails_Johnson_Mittal_Pereira2020}, which uses 2,400 Tor clients and 250 relays. There are significantly more Tor clients in our simulation, as recommended in the latest Shadow study~\cite{Janse_Traudt_Hopper2018}. 



\subsubsection{Simulation Results}

\begin{figure*}[ht]
  \centering
\begin{subfigure}{0.3\textwidth}
\includegraphics[width=\textwidth]{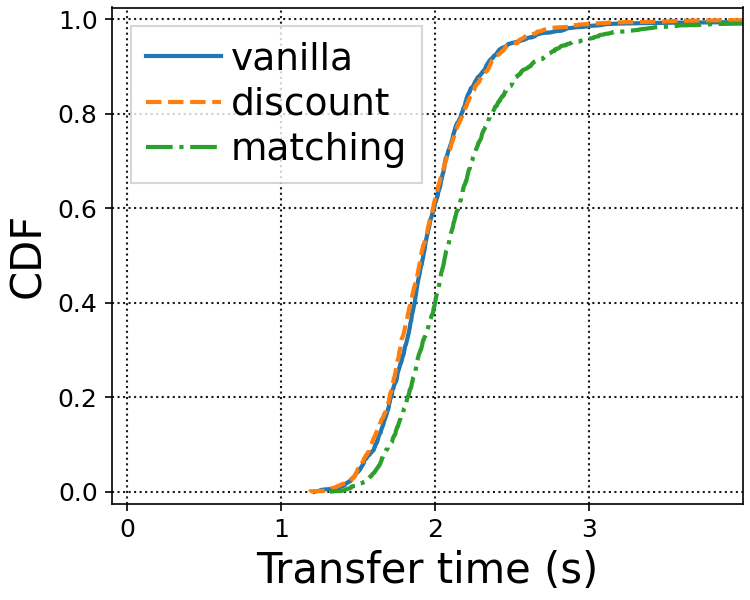}
  \subcaption{Time to last byte received - 1MB}
  \label{9a}
\end{subfigure}
\begin{subfigure}{0.3\textwidth}
\includegraphics[width=\textwidth]{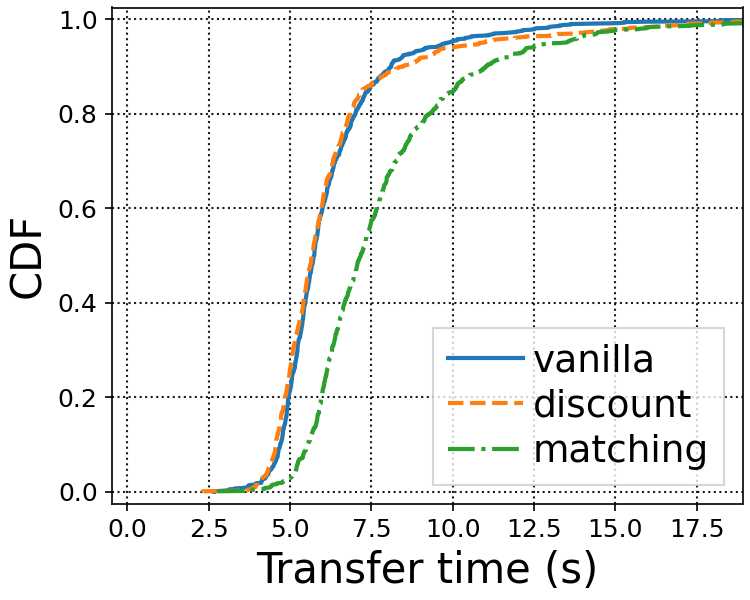}
  \subcaption{Time to last byte received - 10MB}
  \label{9b}
\end{subfigure}
\begin{subfigure}{0.3\textwidth}
\includegraphics[width=\textwidth]{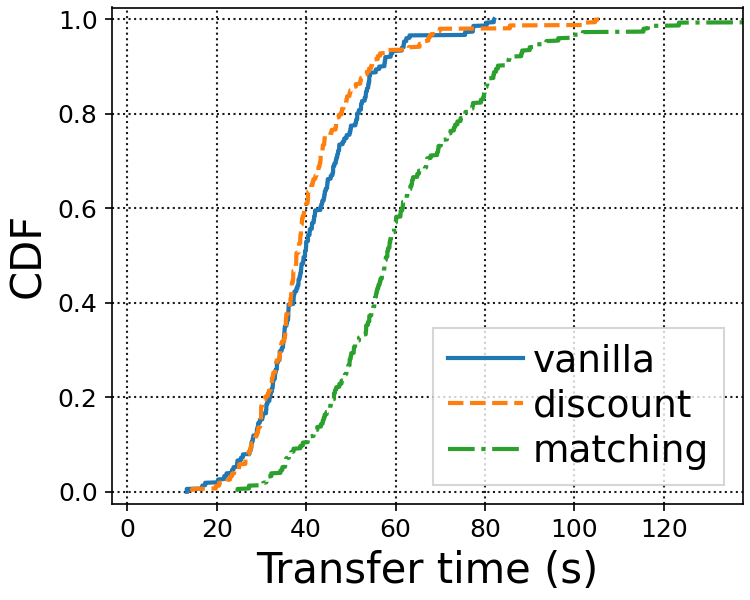}
  \subcaption{Time to last byte received - 100MB}
  \label{9c}
\end{subfigure}
\caption{Simulation running vanilla vs discount vs matching for selected byte size transfers}
\vspace{-5pt}
\label{Matching-Shadow-selected}
\end{figure*}

The most important consideration in performance evaluation is whether running our algorithm will cause any significant performance degradation in latency compared to vanilla Tor. 
Therefore, we focus on the metric time-to-last-byte, i.e., the time it takes between a client initiates a connection and it receives the last byte from the connection. 



Figure \ref{Matching-Shadow-selected} shows the time to last byte received for ten simulations of a 10\% scale of Tor network running the vanilla guard selection, our discount selection algorithm and our matching selection algorithm for selected sizes of data transfer(1MB, 10MB and 100MB). Additional results including 5MB and 50MB transfers are in Appendix \ref{ttlb-complete}.
This result is roughly comparable to the performance reported in CLAPS~\cite{Rochet_Wails_Johnson_Mittal_Pereira2020}, which uses a much smaller simulation scale and shows a data size of 2MB (Figure~\ref{9a} shows 1MB). In particular, our client-relay ratio (100:1) is significantly larger than CLAPS' (10:1). 
For larger data size of 100MB (Figure~\ref{9c}), there is a noticeable gap between vanilla Tor and the matching algorithm, 
where the load time is 6.5 seconds slower on average for the matching algorithm compared to vanilla Tor, which takes an average of 31 seconds. 
However, the median load time for the matching algorithm is 15 seconds slower than vanilla Tor. 
The differences in both average and median load times indicate that for large data loads, the performance is much more volatile for the matching algorithm and the performance degradation may be worse for certain clients. 


On the other hand, there is no noticeable difference between discount and vanilla Tor, even for large transfer sizes of 100MB.

One keynote worth pointing out here is that the results above are purely evaluating time to last byte received for large transfers and is not indicative of the overall network slowdown. As explained in \cite{tmodel-ccs2018}, the probability of large file transfers decreases significantly as transfer size increases. 
After considering the probability of different file size transfers occurring, an average client is only 1.8 seconds slower running our matching selection algorithm compared to vanilla Tor, out of an average of 12 seconds for all file transfers.

Additionally, we evaluate using other metrics such as circuit round trip time, client transfer goodput and relays goodput (see Appendix \ref{Additional_Shadow}). In particular, the results suggest that there is no noticeable difference in client and relay goodput between vanilla Tor and our discount and matching algorithms. For circuit build time and round trip time, there is negligible difference between vanilla Tor and discount selection algorithm and minor difference between vanilla Tor and matching selection algorithm. We find this gap reasonable as matching selection algorithm prefer matched relays, not necessarily the fastest relays. Quantitatively, this gap is 0.057 seconds (5.3\%) slower for circuit build and only 0.0091 seconds (2.4\%)  slower for circuit round trip time.


\textbf{Takeaway.} Our large scale Shadow simulation shows no difference in data transfer time between vanilla Tor and the discount algorithm. There is minimal difference between vanilla Tor and the matching algorithm for smaller data transfer sizes (e.g., 10MB), comparable to algorithms evaluated in the state-of-the-art CLAPS. 
On the other hand, the difference becomes more significant for larger sizes (e.g., 100MB), which usually occur less frequently and have not been evaluated in any prior work.

\section{Deployment Considerations}
\label{sec:deployment}

To deploy either the discount relay selection algorithm or the matching relay selection algorithm, changes need to be made to the Tor network. 
Note that there is no change required for Tor clients in either algorithm, making it easier to deploy. 
All changes will be made to the Tor Directory Authorities. 

For the discount algorithm where relays without ROA will be assigned a discounted weight, the only change that Directory Authorities need to make is to monitor relay ROA status and apply discount to the weight distributed in the consensus accordingly. This is a relatively lightweight and straightforward process. 

The matching algorithm, on the other hand, is more sophisticated and requires additional changes for the Directory Authorities. 
The Directory Authorities need to monitor relays' ROA/ROV status, 
run the weight optimization process, 
and check the client ROA/ROV status and distribute the corresponding set of optimized weights. 
From a client's perspective, it will perform relay selection the same way as vanilla Tor, only using a different set of weights, while the entire optimization process is completely hidden from clients.

\subsection{Client Churn} 
A realistic consideration is the Tor client churn. Every day, tens of thousands of Tor clients \cite{Mani_Brown_Jansen_Johnson_Sherr2018} connect to or drop from the Tor network. 
Such churn may impact the effectiveness of relay selection algorithms that rely on client information (e.g., ROA/ROV status). 
Simulations of static clients performing relay selections may fail to capture this dynamic process. To address this concern, we first quantify the churn on client ROA/ROV distribution---the essential property to the matching algorithm---resulting from client churn on a daily basis.
We then modify our simulations to incorporate such daily changes. 
Note that the discount algorithm is client-agnostic, therefore client churn will not impact the relay selection process. 
Our evaluation below will focus on the matching algorithm, where client churn directly changes the client ROA and ROV distribution and therefore affecting the selection process.

\textbf{Quantifying client ROA/ROV churn.} 
Since we do not have data on hourly influx and outflux of Tor clients, 
we use daily geographic distribution of Tor clients obtained from Tor metrics to model the daily client churn, from January 2024 to April 2024. We choose a four-month period because of the frequency of guard relay selection performed by clients (i.e., every 3-4 months). 
For each day, we compute client ROA and ROV distribution based on the geographic distribution using the same methodology outlined in Section 5.2. 
The results show that client ROV distribution stays steadily at the same level while client ROA distribution exhibits minor fluctuations within the four-month window.
We show the detailed graph in Appendix ~\ref{app:client_churn}.



\begin{figure}[ht]
  \centering
  \includegraphics[scale=0.5]{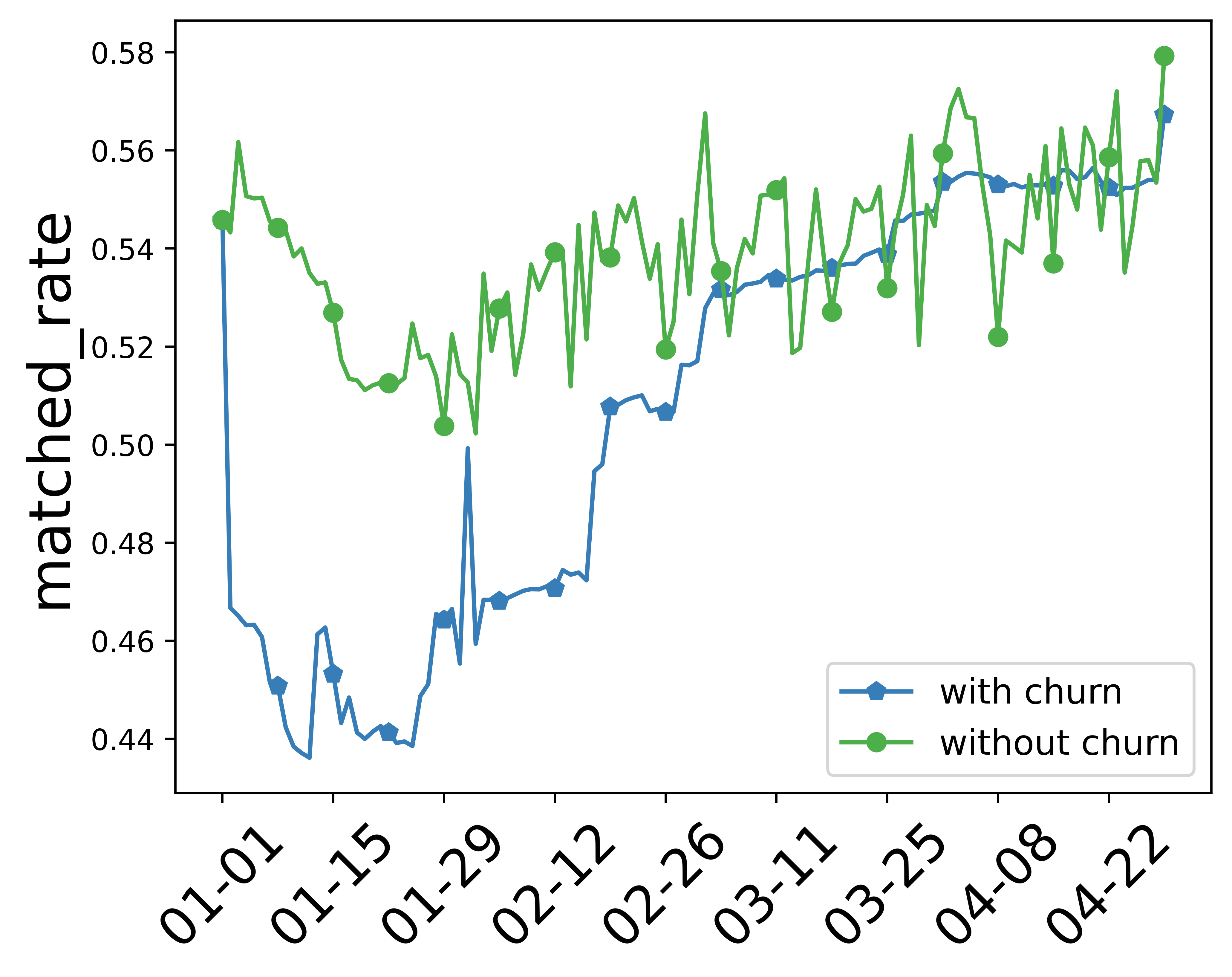}
  \caption{Matched rate with and without churn}
  \vspace{-6pt}
  \label{Churn-Jan-April-2024}
\end{figure}

\textbf{Incorporating client churn into simulation.} We simulate 1 million clients performing relay selection using the Matching Selection algorithm from January 1, 2024 to April 30, 2024, with the impact of client churn. We assume once a client finishes the relay selection process, it will keep using the same guard relay for four months before it performs guard relay selection again, as specified in the current Tor design \cite{arma2013}. 
All clients---generated based on client distribution on Jan 1---will perform guard relay selection using the matching algorithm.
Then for each following day X, we compute the delta between client ROA/ROV distribution on day X and client ROA/ROV distribution on day $X-1$. 
Based on the delta, we generate new clients to categories that increase while removing clients from categories that decrease.
After adjusting the clients, only \emph{new} clients will perform relay selection as existing clients will keep using the guard relays they previously selected. New clients will choose relays based on optimized weights newly computed from the latest consensuses. The dynamic load balancing is in place that records relay selections for clients from the prior days and only load balance new clients to prevent overloading. Figure \ref{Churn-Jan-April-2024} displays the matched rate with and without considering churn.

\textbf{Matched rate with churn is slightly lower than without churn.} From the graph, we can observe that the matched rate with churn is usually lower than without churn.
Depending on the client distribution, difference between the two matched rates averages around 3.15\%, with maximum difference being 10.29\% on January 11 and minimum difference being 0.9\% on March 23.

\textbf{Discussion on ROA/ROV status change for the same client.} Clients may change locations and/or network configurations that results in the change of their AS, which in turn changes their ROA and ROV status. A potential outcome is that the client continues using a guard relay that no longer forms a ROA-ROV client-relay pair. Since our matching selection algorithm requires no modification to the Tor client and all procedures are performed on the directory authorities, we cannot force clients to re-perform the guard selection process and even if we do, this could lead to other security implications. Due to the limited measurement data on Tor clients switching ASes, our simulation could not capture this dynamic. 
We instead focus on measuring global client ROA/ROV distribution changes to gain a high-level understanding of the effect of having  clients entering and leaving Tor. 
In deployment, the client will continue using its previously chosen guard relay until it is time to select another guard, even if the client changes location and loses the protection by using the same guard. This behavior is consistent with the current Tor client. This limitation due to client mobility affects our matching algorithm as well as other relay selection algorithms that rely on client information~\cite{Rochet_Wails_Johnson_Mittal_Pereira2020, Sun_Edmundson_Feamster_Chiang_Mittal2017, Barton_Wright2016}. However, the limitation does not affect our discount relay selection algorithm that is independent of any client information.

{A potential leakage introduced by client churn is when an adversary observes an unmatched client relay pair, she/he may be able to infer (with certain probability) the ROA/ROV status of the previous AS location of the client based on the ROA/ROV status of the currently selected guard relay. Such information is very coarse and does not reveal the specific AS or relay directly. Even though the possibility of an attack using such leakage information may be remote, we address this potential leakage as it is possible that some future changes in the ROA/ROV distributions may result in a very small set of ASes in a given ROA/ROV category, in which case the information leakage increases as the set size decreases.


\subsection{Scalability and Feasibility}
To understand the scalability and feasibility of the relay selection algorithm, we measure the runtime of a single iteration of optimization process on 1 million clients to make sure it is realistic for Directory Authorities to finish the process within a reasonable time. 
The runtime is measured by running Python 3.10 on a Ubuntu 22.04 machine with 8 cores and 16GB of memory. We breakdown all subprocesses by the frequency they should be run with.

Daily routines include updating the ROA and ROV databases, IP to ASN dataset, and client distribution based on Tor monitoring data. The loading time for these datasets in our Python simulation takes about 400 seconds in total. In addition, ROA and ROV checking for the daily client distribution need to be performed, which takes an estimate of 4500 seconds for 1 million clients. Additionally, the ROV database can be updated less frequently given its slow change. It is worth pointing out that the current implementation checks every client's ROA and ROV, which can be further optimized by recording old client ROA/ROV status with churn information to skip partial checking and speedup the process for the next day.

Hourly processes to generate the new consensus weights include calculating relay ROA and ROV distribution based on latest relay information (3.58 seconds) and solving the linear optimization (2.10 seconds). The daily steps take less than 6 seconds in total and may vary slightly based on the ROA/ROV distributions. The linear optimization step is deterministic. As long as the same overall client and relay ROA/ROV distribution is used, it will always produce the same set of updated weights and there will be no conflicts among directory authorities.

\section{Related Work and Discussion}
\label{sec:discussion}

We discuss prior works and the advantages and limitations of our approach. We also discuss potential expansion of our algorithms. 

\subsection{Prior Works}
Many prior works have proposed path selection algorithms for Tor to improve security or performance~\cite{Akhoondi_Yu_Madhyastha2012, Annessi_Schmiedecker2016, Barton_Wright_Ming_Imani2018, Geddes_Schliep_Hopper2016, Panchenko_Lanze_Engel2012, Sherr_Blaze_Loo_Goldberg_Atallah2009, Wacek_Tan_Bauer_Sherr2013, Wang_Bauer_Forero_Goldberg_Keromytis2012, Sun_Edmundson_Feamster_Chiang_Mittal2017, Rochet_Wails_Johnson_Mittal_Pereira2020, Backes_Kate_Meiser_Mohammadi2014, Rochet_Pereira2016, Snader_Borisov2008, Barton_Wright2016, Edman_Syverson2009, Hanley_Sun_Wagh_Mittal2019, Kohls_Jansen_Rupprecht_Holz_Poepper2019, Sun_Edmundson_Feamster_Chiang_Mittal2017, Nithyanand_Starov_Zair_Gill_Schapira_2015, Li_Herwig_Levin2017}.

\textbf{Tor relay selection algorithms against BGP attacks.} 
Counter-RAPTOR~\cite{Sun_Edmundson_Feamster_Chiang_Mittal2017} 
is the first work that defends against active BGP attacks by choosing guard relays located in ASes that are more resilient (i.e., less likely to be affected) by BGP attacks. 
DPSelect~\cite{Hanley_Sun_Wagh_Mittal2019} improves Counter-RAPTOR by employing differential privacy to limit the privacy loss on client location information leakage. 
CLAPS~\cite{Rochet_Wails_Johnson_Mittal_Pereira2020}} improves location-aware relay selection algorithms broadly, including Counter-RAPTOR, by grouping clients by location masks to limit information leakage on client locations. 
Clients obtain relay weights from the group they belong to and perform relay selection as vanilla Tor. One drawback of CLAPS is that relays need to be grouped prior to the linear optimization and this approach adds an extra layer of computation overhead and intransparency to the already complicated relay selection process.

\textbf{Other Tor relay selection algorithms.} 
Path selection algorithms may be location-aware \cite{Barton_Wright2016, Edman_Syverson2009, Hanley_Sun_Wagh_Mittal2019, Kohls_Jansen_Rupprecht_Holz_Poepper2019, Sun_Edmundson_Feamster_Chiang_Mittal2017, Nithyanand_Starov_Zair_Gill_Schapira_2015, Li_Herwig_Levin2017,
Johnson_Jansen_Jaggard_Feigenbaum_Syverson2017} and not location-aware \cite{Backes_Kate_Meiser_Mohammadi2014, Rochet_Pereira2016, Snader_Borisov2008}. 
Not location-aware algorithms run without depending on a client's location and therefore are immune to client location information leakage. 
However, location-aware path selection algorithms, on the other hand, rely on client location and thus leak client information that may lead to deanonymization attacks. They may also be subject to relay placement attacks. Finally, selection algorithms that deviate from bandwidth-only algorithms may be subject to performance downgrade.

\textbf{Defenses against BGP hijacks.} Many works have been done to address BGP hijacks in various directions, 
such as BGP monitoring~\cite{Buhle_Milolidakis_Jacob_Chiesa_Vissicchio_Vanbever2023, Chaviaras_Gigis_Sermpezis_Dimitropoulos2016, Sermpezis_Kotronis_Gigis_Petros_Dimitropoulos_Cicalese_King_Dainotti2018, Obstfeld_Chen_Frebourg_Sudheendra2018}, 
RPKI~\cite{Mirdita_Schulmann_Vogel_Waidner2023, RFC6480, Chung_Aben_Bruijnzeels_Chandrasekaran_Choffnes_Levin_Maggs_Mislove_Rijswijk-Deji_Rula_Sullivan2019, Hlavacek_Cunha_Gilad_Herzberg_Bassett_Schapira_Schulmann2020, Hlavacek_Jeitner_Mirdita_Shulman_Waidner2022_Behind,Hlavacek_Jeitner_Mirdita_Shulman_Waidner2022_Stalloris}, 
and new Internet architecture~\cite{zhang2011scion,scion,birge2022creating}.
These additional secure routing solutions may also be integrated with Tor to provide protections against other forms of routing attacks.

\subsection{Advantages of Location-unaware Relay Selection}
Our discount algorithm is completely independent of any location or other information of the Tor clients. Therefore, it is free of any information leakage resulting from the biases in relay selection. Yet, it still provides a certain level of protection against route origin hijacks. 
On the other hand, although the matching algorithm could potentially leak information on the ROA/ROV status of the Tor client AS, the leakage is minimal compared to location-aware relay selection algorithms even with the clustering approach in CLAPS \cite{Rochet_Wails_Johnson_Mittal_Pereira2020} and the differential privacy approach in DPSelect \cite{Hanley_Sun_Wagh_Mittal2019}. 
Furthermore, despite using a similar optimization framework, the matching algorithm improves upon CLAPS \cite{Rochet_Wails_Johnson_Mittal_Pereira2020} by completely eliminating the clustering step, and takes advantage of the nature of ROA and ROV statuses to spread out the grouping cost into client ROA and ROV checking. By performing simple client ROA and ROV checking, it avoids imposing significant burden on server. Additionally, the Matching algorithm does not require clients to connect to fixed directory authorities like CLAPS do, as each directory authority is capable of perform ROA and ROV checking, which can significantly improve network fault tolerance.

\subsection{Limitations and Future Integration}

While integrating ROA/ROV is a simple yet effective way to defend against route origin hijacks in Tor, it only provides protection against \emph{origin hijacks}, where an AS falsely claims itself as the origin of a prefix. 
It does not prevent routing attacks where the adversary prepend (a path to) the legitimate prefix
owner’s ASN to the announcement, which will bypass the origin check given that the last hop is the legitimate ASN. 
However, such circumvention methods make the AS path longer and less likely to be selected.

\textbf{How the system can be adapted with additional secure routing solutions.}.
While we choose ROA/ROV---given their readily deployed status---to demonstrate the effectiveness of integrating secure routing solutions, additional secure routing solutions may be integrated in the future to prevent routing attacks beyond origin hijacks. We discuss some of the existing defenses and discuss how they can be integrated with our approach to provide additional protections against other routing attacks.
We categorize them into three types: 
(1) BGP monitoring on paths to relays: Prior works have demonstrated the feasibility of monitoring prefixes covering Tor relays~\cite{Sun_Edmundson_Feamster_Chiang_Mittal2017}. We can extend such monitoring to include the last three hops of paths involving Tor prefixes, and adopt the \emph{route age heuristics}~\cite{birge2018bamboozling} to compute how long the path (last three hops) had been seen to the prefix. If the path had never been observed previously, then it is an indicator of potential attack that warrants further investigation. Such live monitoring is helpful in flagging attacks as they occur instead of prevention; 
(2) BGP path validation: Multiple approaches have been proposed to validate the path instead of only the origin, such as BGPSec~\cite{lepinski2017bgpsec} and Path-End Validation~\cite{Cohen_Gilad_Herzberg_Schapira2016}, despite not being deployed yet. Our relay selection algorithm can be adapted to incorporate information from such path-validation mechanisms. For example, the weight computation can factor in whether there exist \emph{Path-end records} (Path-end validation~\cite{Cohen_Gilad_Herzberg_Schapira2016}) from the AS of a relay, or whether the ASes on the path between client and relay perform BGPSec; 
(3) New Internet architecture: Contrary to the inherent routing insecurity in BGP, new Internet architectures such as SCION~\cite{scion} provide routing security by design. SCION has been deployed with ISPs such as the Swiss Secure Finance Network~\cite{ssfn}. As SCION expands its deployment to enable access to more ISPs and coverage of more Tor relays, one factor in assigning weight could be whether the relay and client have access to the SCION network. Meanwhile, techniques such as SBAS~\cite{birge2022creating} can  be utilized to gain access to SCION through tunneling for end hosts that are not part of the SCION network. Note that SCION operates on publicly available federated networks, which alleviates the concern that users may be exposed to a central entity by routing traffic through SCION.  

\section{Conclusion}
In this paper, we develop new Tor guard relay selection algorithms to defend against route origin hijacks by leveraging advances in RPKI. 
By incorporating ROA and ROV information into the relay selection process, our Discount Selection algorithm and Matching Selection algorithm both lead to higher overall ROA and ROV coverage rate for guard relays selected by Tor clients. 
Our approach sheds light on developing location-unaware relay selection algorithms to achieve similar goals as location-aware relay selection algorithms, with the added benefit of avoiding information leakage on client locations.

\begin{acks}
We thank anonymous reviewers for their insightful and constructive suggestions and feedback. This work is supported by National
Science Foundation CNS-2154962 and CNS-2319421, and the Commonwealth Cyber Initiative.
\end{acks}

\bibliographystyle{ACM-Reference-Format}
\bibliography{sample-base}

\begin{appendices}
\newpage
\section{ROA Coverage for All Tor Relays}
\label{ROA_coverage_all}
This appendix shows the ROA coverage information for all Tor relays for the period between January 2021 and May 2024.
\begin{figure}[h]
  \centering
  \includegraphics[width=\linewidth]{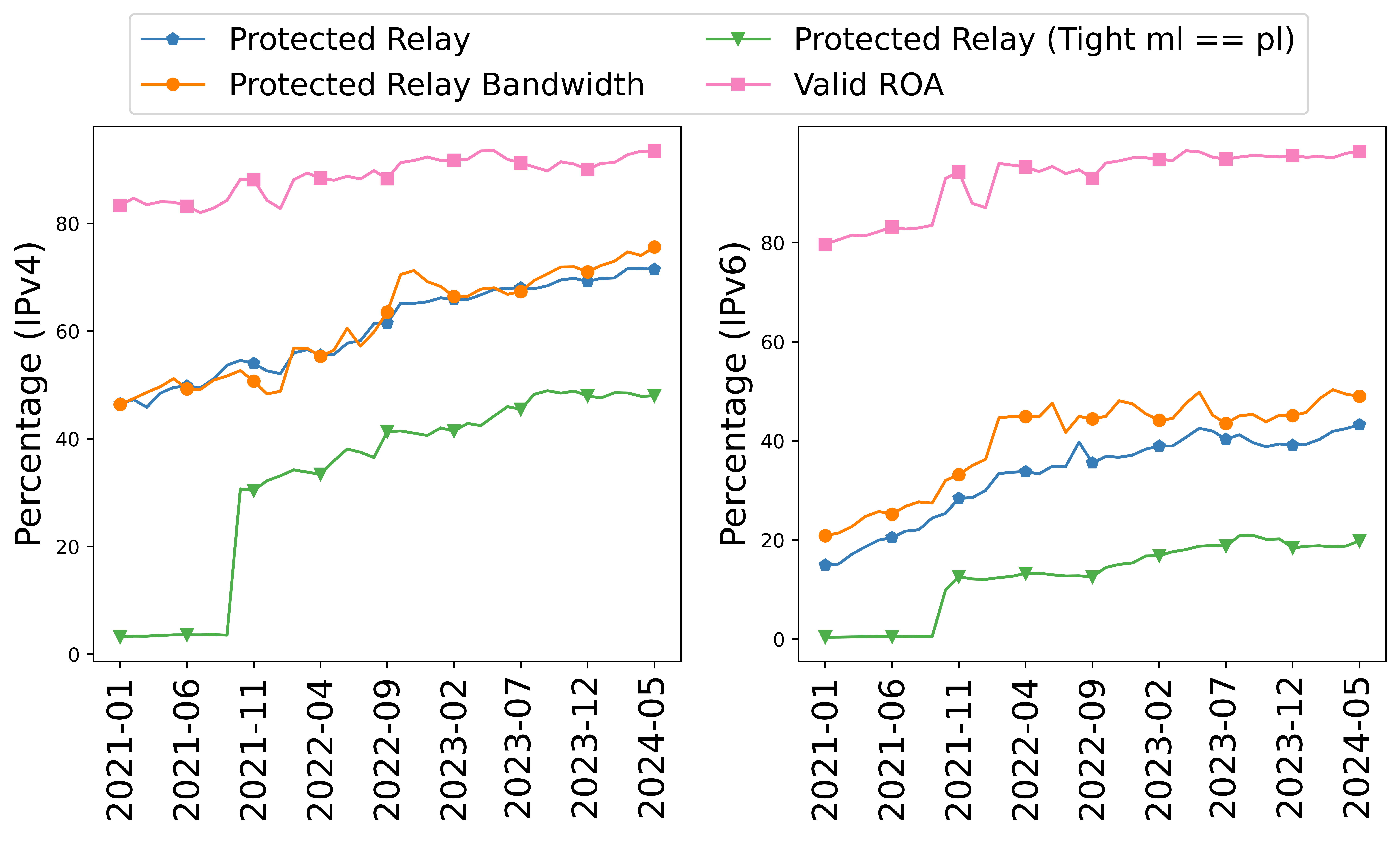}
  \caption{ROA Coverage and Validity for All Tor Relays}
  \label{ROA-All}
\end{figure}

\section{Matching performance under different parameter combinations}
\label{Appendix_A}

This table displays the matched rate and improvement (compared to vanilla Tor) under various parameter combinations.
\begin{table}[h!]
\centering
 \begin{tabular}{||c c c c c c||} 
 \hline
 $l$ & $d_{1}$ & $d_{2}$ & $B$  & Matched Rate & $\Delta$ Matched Rate\\ [1ex] 
 \hline\hline
 0.8 & 0.9 & 0.8 & 1.5 & 0.5619 & 0.4022\\ 
 0.8 & 0.9 & 0.7 & 1.5 & 0.5621 & 0.4022\\
 0.8 & 0.9 & 0.6 & 1.5 & 0.5422 & 0.3826\\
 0.8 & 0.8 & 0.7 & 1.5 & 0.5613 & 0.4013\\
 0.8 & 0.8 & 0.6 & 1.5 & 0.5618 & 0.4019\\
 0.8 & 0.7 & 0.6 & 1.5 & 0.5618 & 0.4025\\ 
 0.6 & 0.9 & 0.7 & 1.5 & 0.6845 & 0.5248\\ 
 0.6 & 0.8 & 0.6 & 1.5 & 0.6841 & 0.5247\\
 0.8 & 0.9 & 0.8 & 2.0 & 0.5616 & 0.4016\\
 0.8 & 0.9 & 0.7 & 2.0 & 0.5615 & 0.4018\\
 0.8 & 0.8 & 0.7 & 2.0 & 0.5618 & 0.4023\\
 0.8 & 0.8 & 0.6 & 2.0 & 0.5619 & 0.4021\\ [1ex]
 \hline
 \end{tabular}
\end{table}

\section{Client ROA/ROV Churn}
\label{app:client_churn}
This appendix displays the client churn in terms of ROA and ROV distribution for the period from January 1, 2024 to April 30, 2024.

\begin{figure}[H]
  \centering
  \includegraphics[scale=0.5]{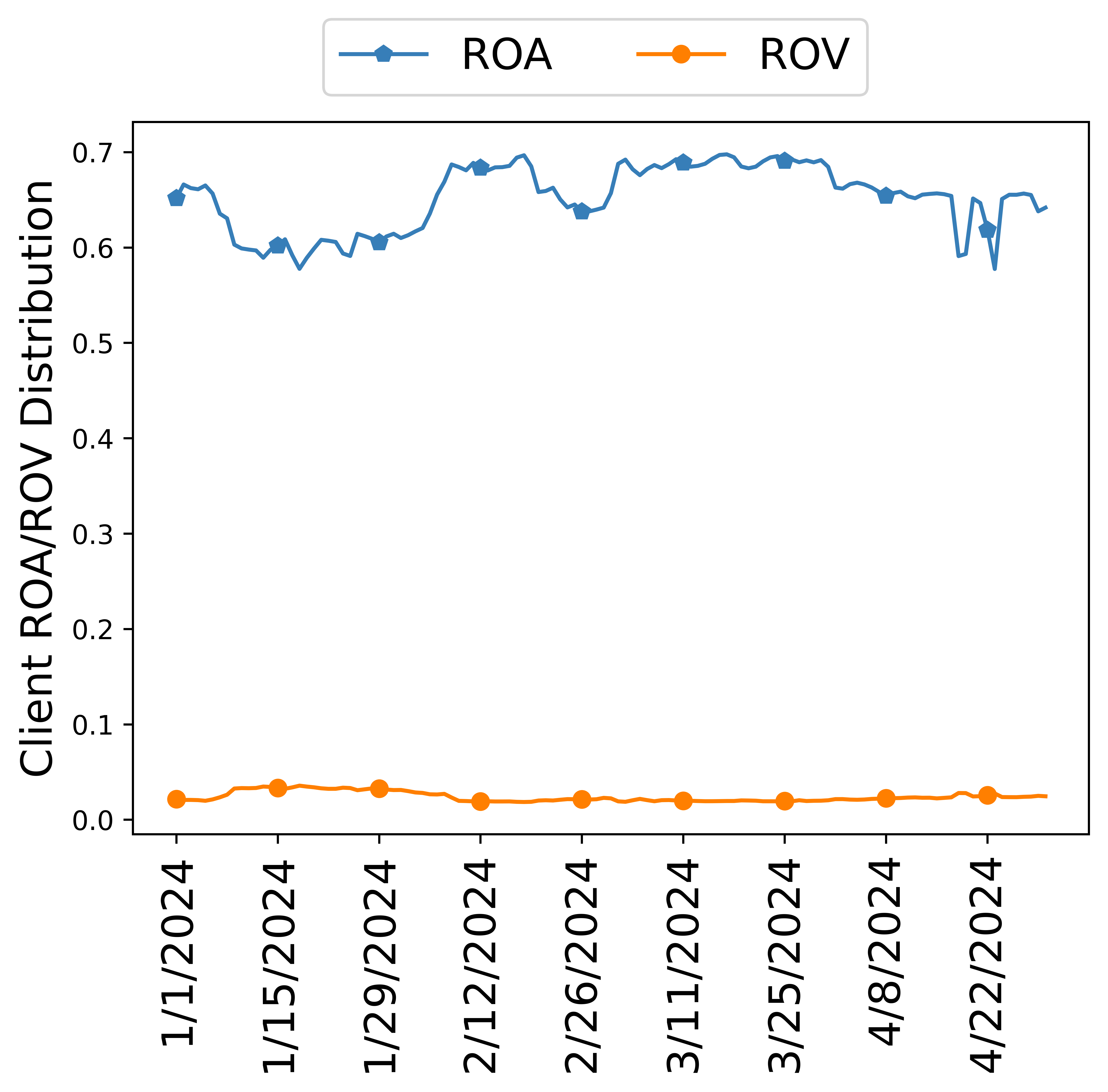}
  \caption{Clients ROA \& ROV distribution}
  \label{ROA-ROV-Jan-April-2024}
\end{figure}

\section{Bandwidth utilization under different discount and load}
\label{Load_x_discount}
This appendix displays the expected actual bandwidth utilization under various combinations of discount and load factors. The expected bandwidth utilization is calculated as the percentage of the sum of all updated consensus weights (after discount) for selected relays out of the sum of original weights for all relays. Therefore, the baseline (when $d=1$) is not a straight 45 degree line, instead it is a skewed line starting at the point where initial load and actual load are both 1 with the slope being the percentage of sum of bandwidths of those relays with ROA. It is important to point out that the x-label is discount factor, while each line represents an inherent initial load factor.
\begin{figure}[H]
  \centering
  \includegraphics[scale=0.5]{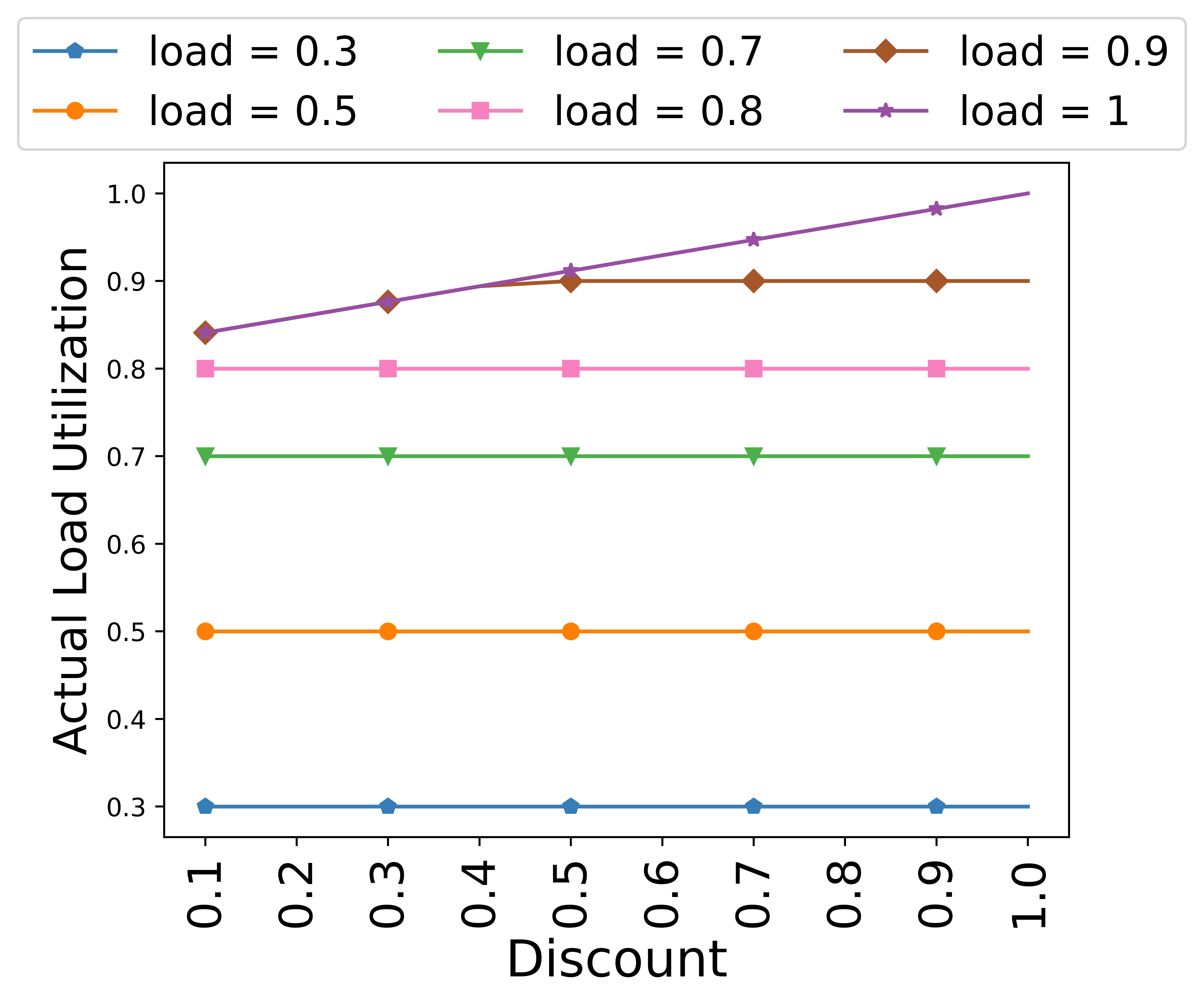}
  \caption{Expected bandwidth utilization given specific load and discount factors, using consensus from May 01, 2024}
  \label{Discount-Load}
\end{figure}

\section{Additional Shadow simulation results}
\label{Additional_Shadow}
This appendix displays various metrics for evaluating Shadow simulations using various guard relay selection methods. The metrics included here are circuit round trip time, circuit build time, relay goodput and client goodput.
\begin{figure}[H]
  \centering
  \includegraphics[scale=0.5]{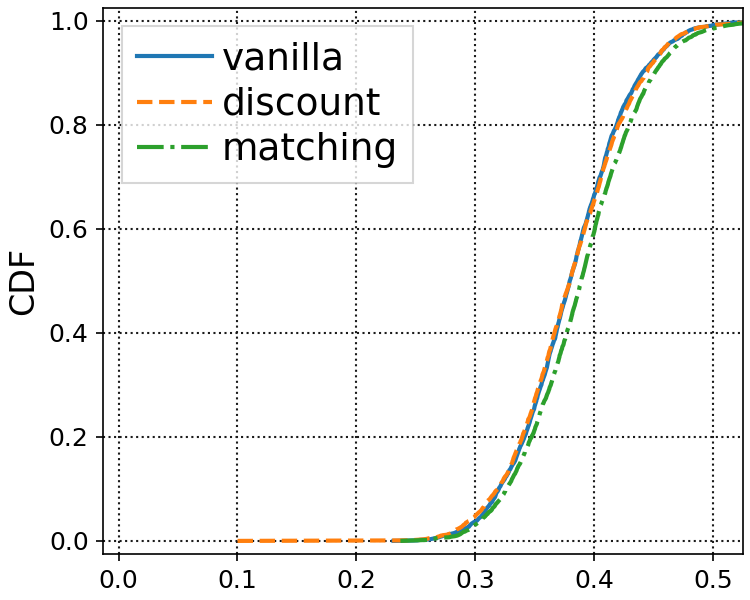}
  \caption{Circuit round trip time (s)}
  \label{Circuit-roundtrip}
\end{figure}

\begin{figure}[H]
  \centering
  \includegraphics[scale=0.5]{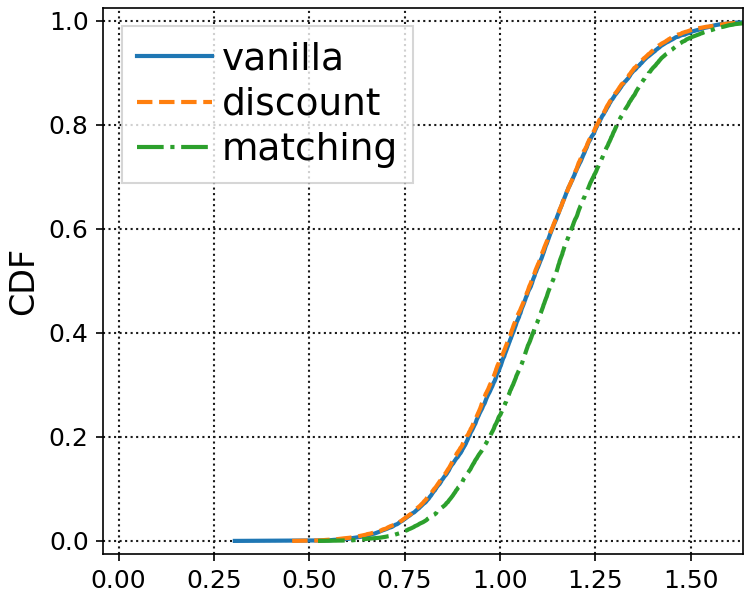}
  \caption{Circuit build time (s)}
  \label{Circuit-build-time}
\end{figure}

\begin{figure}[H]
  \centering
  \includegraphics[scale=0.5]{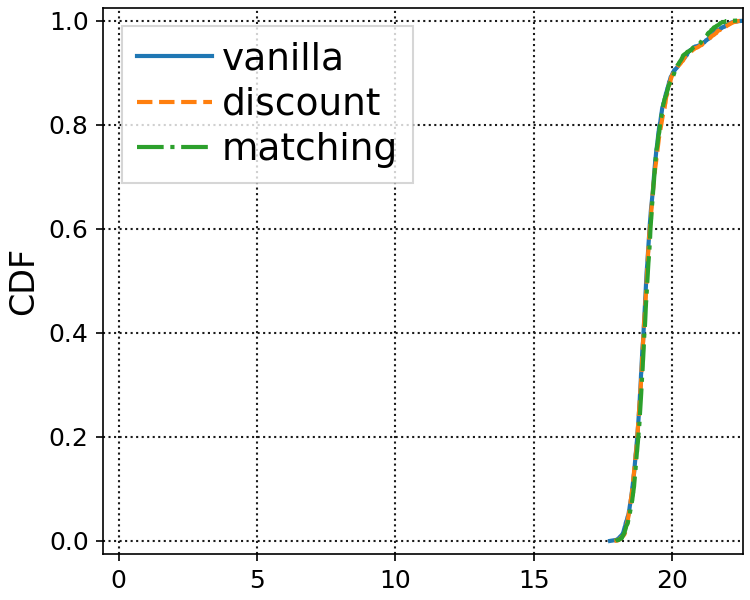}
  \caption{Relay goodput (Gbit/s)}
  \label{Relay-goodput}
\end{figure}

\begin{figure}[H]
  \centering
  \includegraphics[scale=0.5]{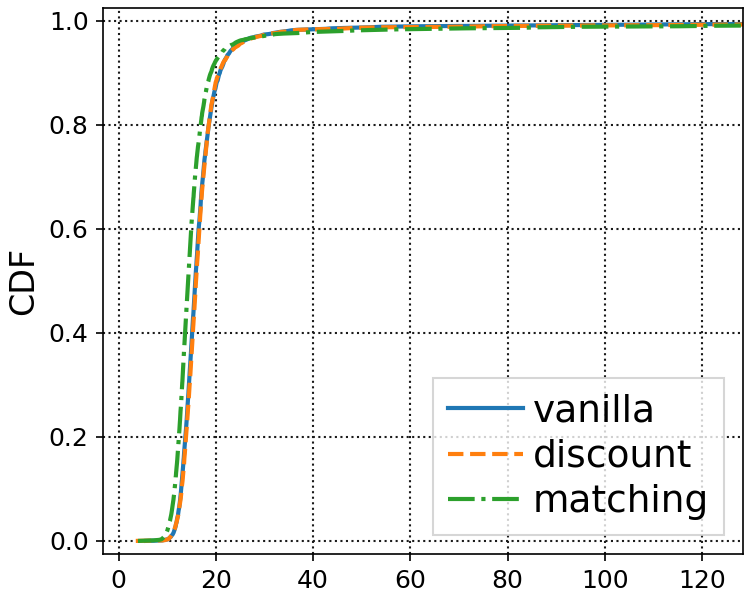}
  \caption{Client goodput (Mbit/s)}
  \label{Client-goodput}
\end{figure}

\section{Time to last byte received for all file transfers}
\label{ttlb-complete}
This appendix displays the last byte received for all byte size transfers, including 1MB, 5MB, 10MB, 50MB, 100MB.
\begin{figure*}[ht]
  \centering
\begin{subfigure}{0.3\textwidth}
\includegraphics[width=\textwidth]{figures/transfer_time_1048576.exit.png}
  \subcaption{Time to last byte received - 1MB}
  \label{14a}
\end{subfigure}
\begin{subfigure}{0.3\textwidth}
\includegraphics[width=\textwidth]{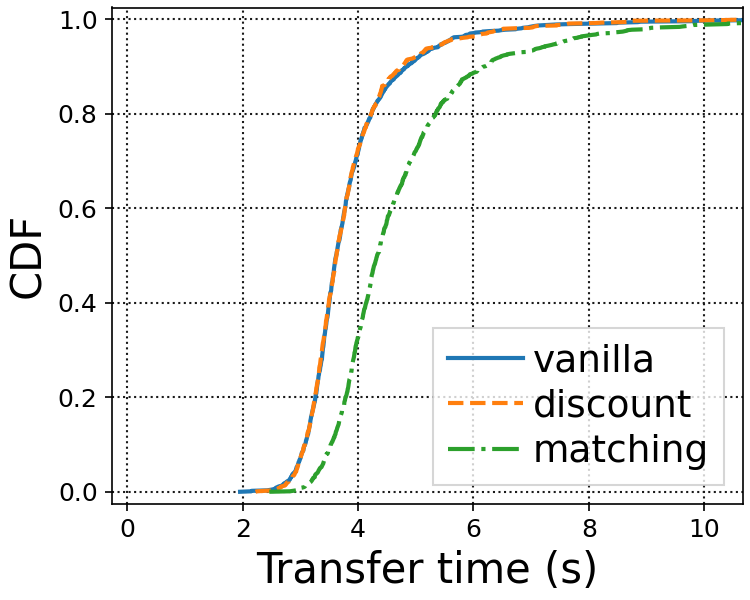}
  \subcaption{Time to last byte received - 5MB}
  \label{14b}
\end{subfigure}
\begin{subfigure}{0.3\textwidth}
\includegraphics[width=\textwidth]{figures/transfer_time_10485760.exit.png}
  \subcaption{Time to last byte received - 10MB}
  \label{14c}
\end{subfigure}
\begin{subfigure}{0.3\textwidth}
\includegraphics[width=\textwidth]{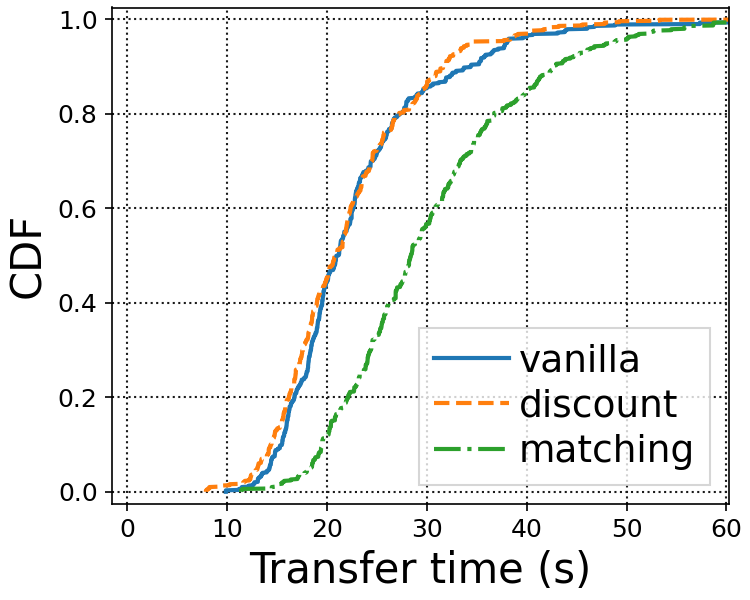}
  \subcaption{Time to last byte received - 50MB}
  \label{14d}
\end{subfigure}
\begin{subfigure}{0.3\textwidth}
\includegraphics[width=\textwidth]{figures/transfer_time_104857600.exit.png}
  \subcaption{Time to last byte received - 100MB}
  \label{14e}
\end{subfigure}
\caption{Simulation running vanilla vs discount vs matching for different byte size transfers}
\label{Matching-Shadow}
\end{figure*}
\end{appendices}

\end{document}